\documentclass[11pt,preprint]{emulateapj}
\usepackage{lscape,url,color}
\usepackage{amsmath}
\usepackage{relsize}

\newcommand{\logg}{\mbox{$\log g$}}
\newcommand{\teff}{\mbox{T$_\mathrm{eff}$}}
\newcommand{\fsed}{\mbox{$f_\mathrm{sed}$}}

\newcommand{\mjup}{M$_{\rm Jup}$}

\newcommand{\MJ}{M$_{J}$}
\newcommand{\MWT}{M$_{W2}$}

\newcommand{\mko}{$_{MKO}$}

\shortauthors{Tinney {\it et~al.\/}}
\shorttitle{Luminosities of the Coldest Brown Dwarfs}
\slugcomment{To appear in The Astrophysical Journal}
\received{8 July, 2014}
\accepted{2 October, 2014}
\begin{document}

\title{The Luminosities of the Coldest Brown Dwarfs\altaffilmark{*}}

\author{
C. G. Tinney\altaffilmark{a,b},
Jacqueline K. Faherty\altaffilmark{c},
J. Davy Kirkpatrick\altaffilmark{d}, 
Mike Cushing\altaffilmark{e},
Caroline V. Morley\altaffilmark{f},
Edward L. Wright\altaffilmark{g}
}

\altaffiltext{*}{This paper includes data gathered with the 6.5 meter Magellan Telescopes located at Las Campanas Observatory, Chile.}
\altaffiltext{a}{School of Physics, UNSW Australia, NSW 2052, Australia. c.tinney@unsw.edu.au}
\altaffiltext{b}{Australian Centre for Astrobiology, UNSW Australia, NSW 2052, Australia}
\altaffiltext{c}{Department of Terrestrial Magnetism, Carnegie Institution of Washington, Washington DC, USA}
\altaffiltext{d}{Infrared Processing and Analysis Center, MS100-22, California Institute of Technology, Pasadena, CA 91125, USA}
\altaffiltext{e}{Department of Physics and Astronomy, The University of Toledo, OH 43606, USA }
\altaffiltext{f}{Department of Astronomy and Astrophysics, University of California, Santa Cruz, CA 95064. USA}
\altaffiltext{g}{Department of Physics and Astronomy, UCLA, Los Angeles, CA 90095-1547, USA}


\begin{abstract}
In recent years brown dwarfs have been extended to a new
Y-dwarf class with effective temperatures colder than 500K and masses in the range 5-30 Jupiter masses. 
They fill a crucial gap in observable atmospheric properties between the  much 
colder gas-giant planets of our own Solar System (at around 130K) and both hotter T-type brown dwarfs
and the hotter planets that can be imaged orbiting young nearby stars (both with effective temperatures of in the range 1500-1000K).
Distance measurements for these objects 
deliver absolute magnitudes that make critical tests of our understanding of very cool atmospheres. Here 
we report new distances for nine Y dwarfs and seven very-late T dwarfs. These reveal that Y dwarfs
do indeed represent a continuation of the T dwarf sequence to both fainter luminosities and cooler
temperatures. 
They also show that the coolest objects display a large range in absolute magnitude for a given
photometric colour. The latest atmospheric models  show good agreement with the majority of 
these Y dwarf absolute magnitudes. This is also the case for WISE0855-0714 the coldest and closest brown dwarf 
to the Sun, which shows evidence for water ice clouds. However, there are also some outstanding
exceptions, which suggest either binarity or the presence of condensate clouds. The former is readily 
testable with current adaptive optics facilities. The latter would mean that the range of cloudiness 
in Y dwarfs is substantial with most hosting almost no clouds -- while others have dense clouds making 
them prime targets for future variability observations to study cloud dynamics.
\end{abstract}

\section{Introduction} \label{intro}

Y-type brown dwarfs were first discovered by the WISE satellite \citep{Cushing11} as an extension to colder temperatures of the L-
and T-type sequences for stars and brown dwarfs\footnote{We note also the detection of the companion to the white dwarf 
WD\,0806-661 by \citet{Luhman11}, and the detection of a faint companion in the CFBDSIR\,J1458+1013AB system \citep{Liu11b}. 
Though both these objects\citep[like W0855 discovered by][]{Luhman14} are too faint to allow an optical or near-infrared spectrum to be obtained, 
they are very likely Y dwarfs.}  \citep{ReidHawley06}. Interpretation of the physical properties of these unusual objects
has been primarily driven by modelling of their observed spectra \citep{Kirkpatrick13a,Cushing11,Leggett13} 
using a variety of atmospheric models \citep{Allard03,SaumonMarley08,Morley14}.
However, while the interior structures for brown dwarfs are relatively straightforward \citep{Stevenson91}, atmospheres represent a
substantial challenge. At these very cold temperatures a rich panoply of molecular physics in a wide range of
species (including CO, CH$_4$, N2, NH3, H$_2$O, Fe, Cr, CaTiO3, Na$_2$S, MnS, ZnS and KCl) becomes critical -- not only for
modelling the opacity as a function of wavelength through the atmosphere, but also for determining the physical
structure of the atmosphere (i.e. whether condensates are present), as well as the atmosphere's chemical make-up. Since condensed
molecules are removed from the gas-phase chemistry, their absorption opacities are removed from the radiative
transfer, but have to be accounted for in the form of clouds of scatterers and absorbers. The physical height at
which cloud layers of various materials will settle (as a function of each object's overall effective temperature
and specific atmospheric structure) therefore becomes a vital component of atmospheric models.

Examples of the impact of such cloud formation on the emergent spectra of brown dwarfs include: the evolution of observable molecular
bands of TiO, VO and FeH through the L dwarf sequence \citep{Tsuji00}; the transition from cloudy late L-dwarfs to relatively
cloudless early T dwarfs \citep[the specific mechanism for which is still hotly debated, and which allows significant
rotationally-driven variability at this transition][]{SaumonMarley08,Radigan12,Artigau09}; and the associated increase in the J-band absolute
magnitude (the ``J-band hump'') seen in early T dwarfs \citep{Tinney03}.

Distance measurements via trigonometric parallaxes have been key in determining the physics required in succeeding
generations of models, since they provide absolute magnitudes (in a given bandpass), which can be integrated to
determine bolometric luminosities. They are also key for completing the census of objects near the Sun -- as
spectacularly demonstrated by the recent detection of a ~250K brown dwarf at a distance of just 2.2\,pc \citep{Luhman14}. 

The astrometric motions produced by trigonometric parallax  are small -- 40 milliarcseconds (mas) in amplitude for an object
at 25\,pc. However, techniques have advanced tremendously over the last two decades,  and these observations are now
possible with ground-based telescopes in both the optical \citep{Monet92,Dahn02,Tinney95}  and near-infrared \citep{Tinney03,Vrba04,Faherty12,DL12}, 
as well as using space-based facilities like {\em Spitzer} and HST
 \citep{DK13,Beichman14}. 

The {\em Gaia}\footnote{\url{http://sci.esa.int/gaia/}} mission is set to deliver an unprecedented flood of new astrometry 
at optical wavelengths over the next few years.
However for the coldest brown dwarfs, the combination of their faintness and their  flux primarily emerging in the near-to-mid-infrared, 
means that their parallaxes ill remain the exclusive domain of targeted astrometric programs \citep{smart14}.
In 2012 we therefore commenced a new astrometric program
using the  FourStar imaging camera on the 6.5m Magellan Baade telescope to target parallax measurements for very faint
(J $\sim$ 19.5-22) Y dwarfs.

\section{Observations \& Analysis}

Parallax observations were obtained using the FourStar imaging camera \citep{persson2008} on the Magellan Baade telescope between March
10, 2012 and June 18, 2014. FourStar is a near-infrared mosaic imager with four 2048$\times$2048 pixel detectors giving an imaging field of view of
11$\arcmin$\ on a side at a pixel scale of 0.159$\arcsec$/pixel. It is equipped with a set of intermediate-band filters (originally specified
for the measurement of photometric redshifts) which turn out to be almost ideally suited for observing very cool brown dwarfs \citep[see Fig.1 in][]{Tinney12}.
In particular,  the J3 filter ($\lambda_{cen} \approx 1.29\,\mu$m) collects almost all of the J-band flux from late-T and Y dwarfs, 
while only collecting roughly half the night-sky emission of the J-band. All imaging results in this paper were obtained in the FourStar J3 filter.

Image quality over the course of the program varied between 0.29$\arcsec$ and 1.19$\arcsec$ with a
median of 0.61$\arcsec$ (see Table \ref{Runs}).  Our astrometric observing and analysis techniques follow those
previously described by us \citep{Tinney12,Tinney95}, and involve observing each target with the FourStar J3 filter in a sequence of 60-120s randomly dithered exposures
at very similar hour angles on every night. Targets are observed reaching net
integration times ranging from 15 minutes (for the brightest J$\sim$19 targets) to 1.5-2.0h (for the faintest J$\sim$22 targets).
Images are dark subtracted, flat-fielded and mosaic-combined in a standard manner using a modified version of the
ORACDR\footnote{\url{http://www.jach.hawaii.edu/JACpublic/UKIRT/software/oracdr}} data reduction pipeline. This is run twice on each field. 
Individual images revealed to have significantly poorer image quality than the rest of the jittered data set after the first-pass processing 
are removed from the list of images used in the second-pass.

\begin{deluxetable}{lclc}
\tablecaption{Observation epochs at the Magellan Baade Telescope.\label{Runs}}
\tablewidth{0pt}
\tablehead{
\colhead{UT Date}          & \colhead{Median }  & \colhead{UT Date}          & \colhead{Median }  \\
                           & \colhead{Seeing ($\arcsec$)}    &            & \colhead{Seeing ($\arcsec$)}
}
\startdata
2012 Mar 10	& 0.64 & 2013 Jul 27 & 0.81 \\
2012 May 10	& 0.56 & 2013 Aug 15 & 0.43 \\
2012 Jun 6	& 0.70 & 2013 Oct 20 & 0.84 \\
2012 Jun 7	& 0.53 & 2013 Nov 14 & 0.69 \\
2012 Aug 10	& 0.87 & 2013 Dec 11 & 0.51 \\
2012 Oct 6	& 0.82 & 2014 Jan 13 & 0.74 \\
2013 Jan 15	& 0.34 & 2014 Mar 10 & 0.50 \\
2013 Feb 2	& 0.92 & 2014 May 13 & 0.58 \\
2013 Mar 22	& 0.32 & 2014 Jun 17 &  0.53    \\
2013 Apr 22	& 0.54 & 2014 Jun 18 &  0.54    \\
\enddata
\end{deluxetable}

\subsection{Photometric Analysis}
\label{PhotAnal}

Photometry for all our targets was measured using Sextractor aperture photometry \citep{bertin96}. 
Each image has been zero-point calibrated by using 2MASS Point Source Catalogue \citep{skrutskie2006} J-band data in that field to
generate a photometric zero point on the J3 photometric system (see below for details). The multiple epoch observations generate multiple J3
photometric estimates for each target, which have been averaged (weighted by estimated photometric uncertainties) to
generate J3 photometry for our targets which we report in Table \ref{Results} (where the uncertainties quoted
are the resulting standard error in the mean).
The table also presents literature J\mko\  photometry for each target where available, 
along with the full {\em AllWISE} designations and W2 photometry of each target \citep{Wright10,Mainzer11}.

\begin{figure} 
   \figurenum{1}
   \begin{center}
      \includegraphics[trim=10mm 90mm 35mm 120mm, clip=true,width=90mm]{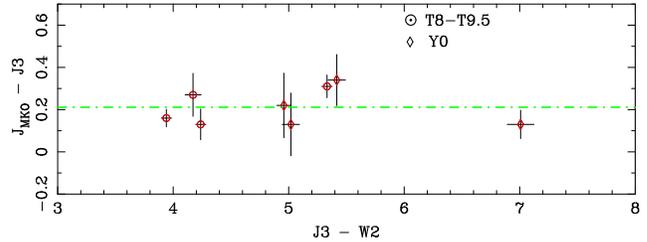}
   \end{center}
   \caption{J\mko--J3 as a function of J3--W2 colour. T dwarfs and Y dwarfs are plotted with distinct symbols, and
   the mean value of J\mko--J3 is indicated by the dot-dashed line.\label{Fig3}}
\end{figure}

Because the J3 filter collects essentially {\em all} of the flux emitted in the
J band for these late T and Y dwarfs, J3 magnitudes can be readily converted to J\mko\ magnitudes. 
We  have used our new J3 data to update the estimate for this correction presented by \citet{Tinney12}.
In Figure \ref{Fig3} we plot J\mko--J3 as a function of J3--W2 for the nine objects in Table \ref{Results} with
J\mko\ measured to better than 0.2\,mag precision. T and Y dwarfs are plotted with different symbols, and the 
mean value of J\mko--J3 = 0.20$\pm$0.03 (where the uncertainty is the standard error in the mean) is plotted as a 
dot-dashed line. The data display no evidence for a significant systematic trend (i.e. greater than 0.1\,mag in size) over
the colour range of interest for Y dwarfs and the latest T dwarfs.

Our J3 data generally has much better
photometric precision than the extant J\mko\  photometry for these objects, and is derived from multiple observations, 
so we use this corrected J3 photometry to generate  
``synthetic'' J\mko\  photometry for all our Magellan targets, and  we use this photometry
whenever we refer to J\mko\ for those objects in the rest of this paper.

\subsection{Astrometric Analysis}

Astrometry of our targets was performed following the procedures adopted in earlier astrometric papers by our 
team \citep{Tinney12,Tinney03,Tinney95}. Briefly, astrometry was measured using the DAOPHOT II package \citep{Stetson87} as implemented within the Starlink
environment provided by the Joint Astronomy Centre, Hawaii\footnote{\url{http://starlink.jach.hawaii.edu/starlink}}.
Each epoch is transferred to a master frame using a ``solid body'' linear transformation that allows for field rotation, plate scale change and field offset (but {\em not} field shear). The master frame is rotated to the cardinal directions and its plate scale is determined using 2MASS reference stars in the field. Typical uncertainties on the plate scale for the master frame are better than $\pm$0.2\% making this a negligible source of uncertainty for the very nearby stars targeted in this program. Astrometric solutions are derived using
codes previously developed by us, and making use of the NOVAS library\footnote{\url{http://aa.usno.navy.mil/software/novas/novas\_info.php}} provided by the US Naval Observatory. 

One substantial difference between these observations and those used in our previous parallax programs is that our
targets are often substantially fainter than our reference frame stars. As such, using the residuals about each
transformation as an estimate for the astrometric uncertainty in each frame (as done by us in the past) does not
correctly reflect the full systematic plus photon-counting uncertainties associated with each frame. We therefore
carried out an additional analysis using the residuals for all objects present in all frames for a given target
(including star-like objects much fainter than the astrometric reference stars). Using the residuals for all these
``pseudo-reference'' stars, binned as a function of magnitude, we were able to fit a model for the astrometric
uncertainty as a function of J3 magnitude in each frame. This model assumes the sky noise over the object photon-counting
noise for each target and is of the form

$$\mathrm{RMS}^2 = a_0^2 ( 1 + 10^{0.8(J3-a1)})$$

Where J3 is the object magnitude, and $a_0$ and $a_1$ are fitting constants, such that $a_0$ parametrises the
underlying systematic residuals associated with each frame's transformation, while $a_1$ parametrises the precision as
a function of magnitude (it is formally the magnitude at which the photon-counting contribution to the precision is
equal to the underlying precision of the reference frame transformation $a_0$). 
This model was then used to estimate the astrometric precision for our target
object in each frame, and these uncertainties are used in the parallax solution.

\section{Results}

\begin{figure*} 
   \figurenum{2}
   \begin{center}
      \includegraphics[trim=5mm 75mm 25mm 15mm, clip=true,width=65mm]{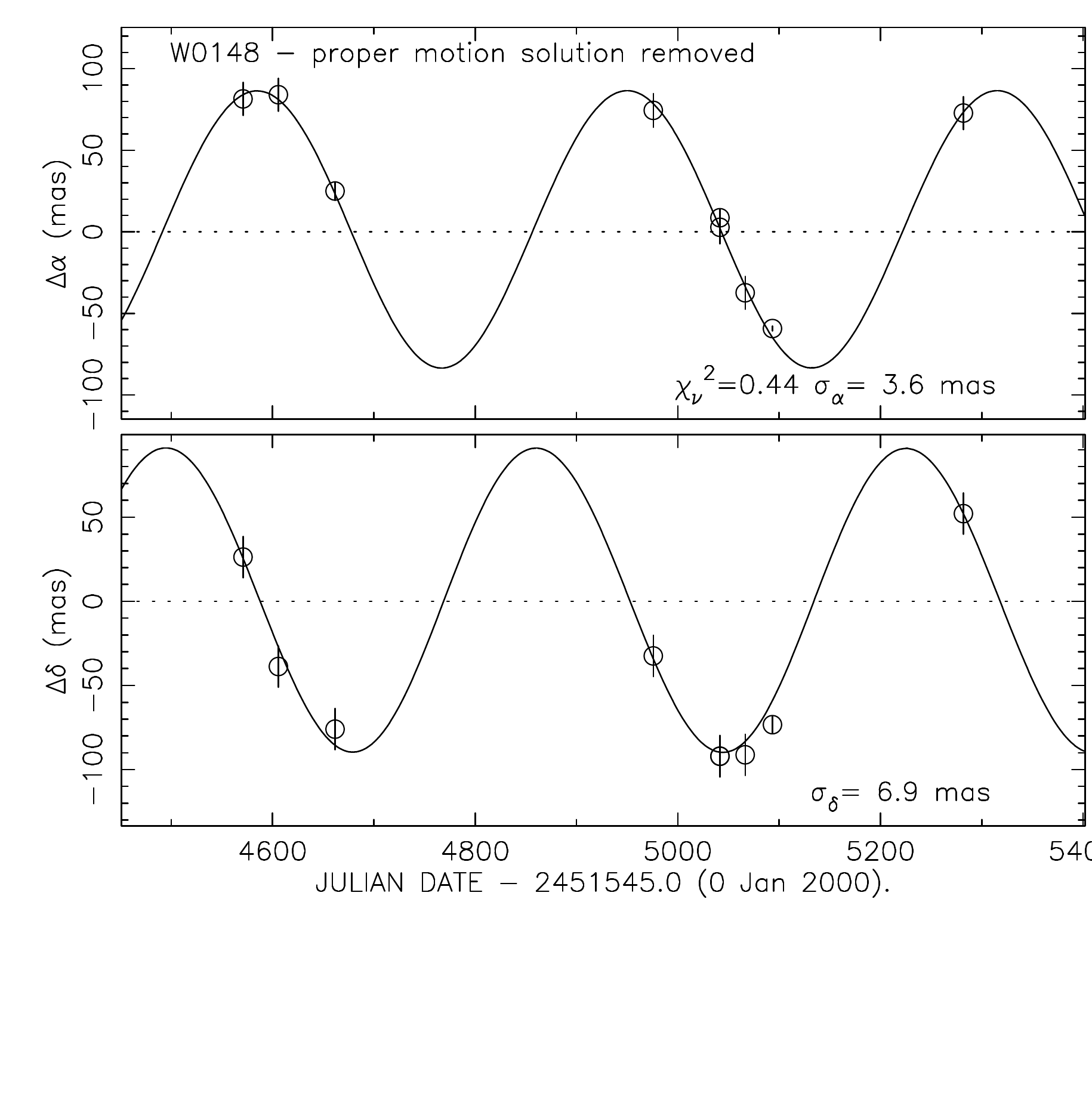}
      \includegraphics[trim=5mm 75mm 25mm 15mm, clip=true,width=65mm]{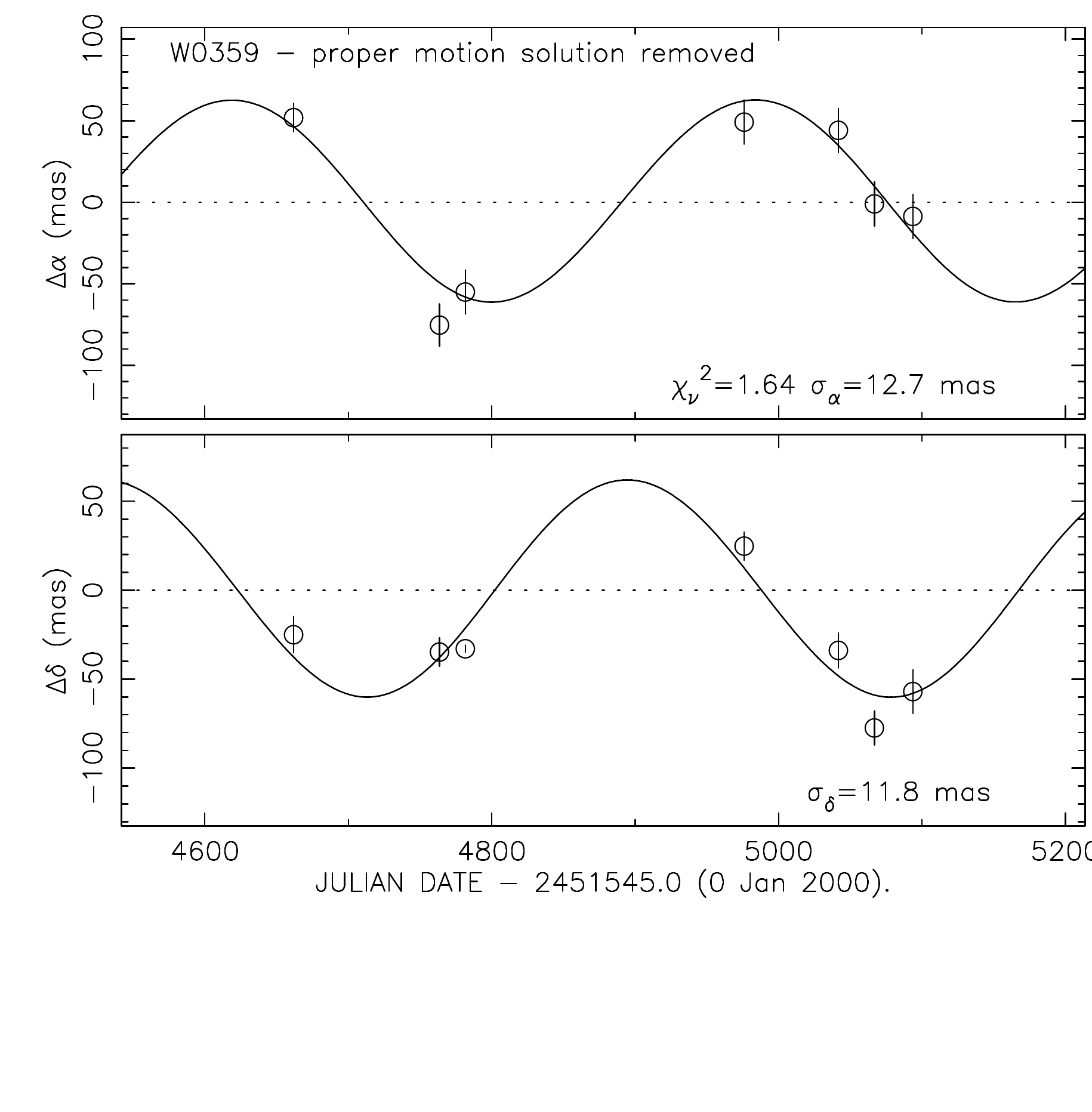}\\[-35pt]
      \includegraphics[trim=5mm 75mm 25mm 15mm, clip=true,width=65mm]{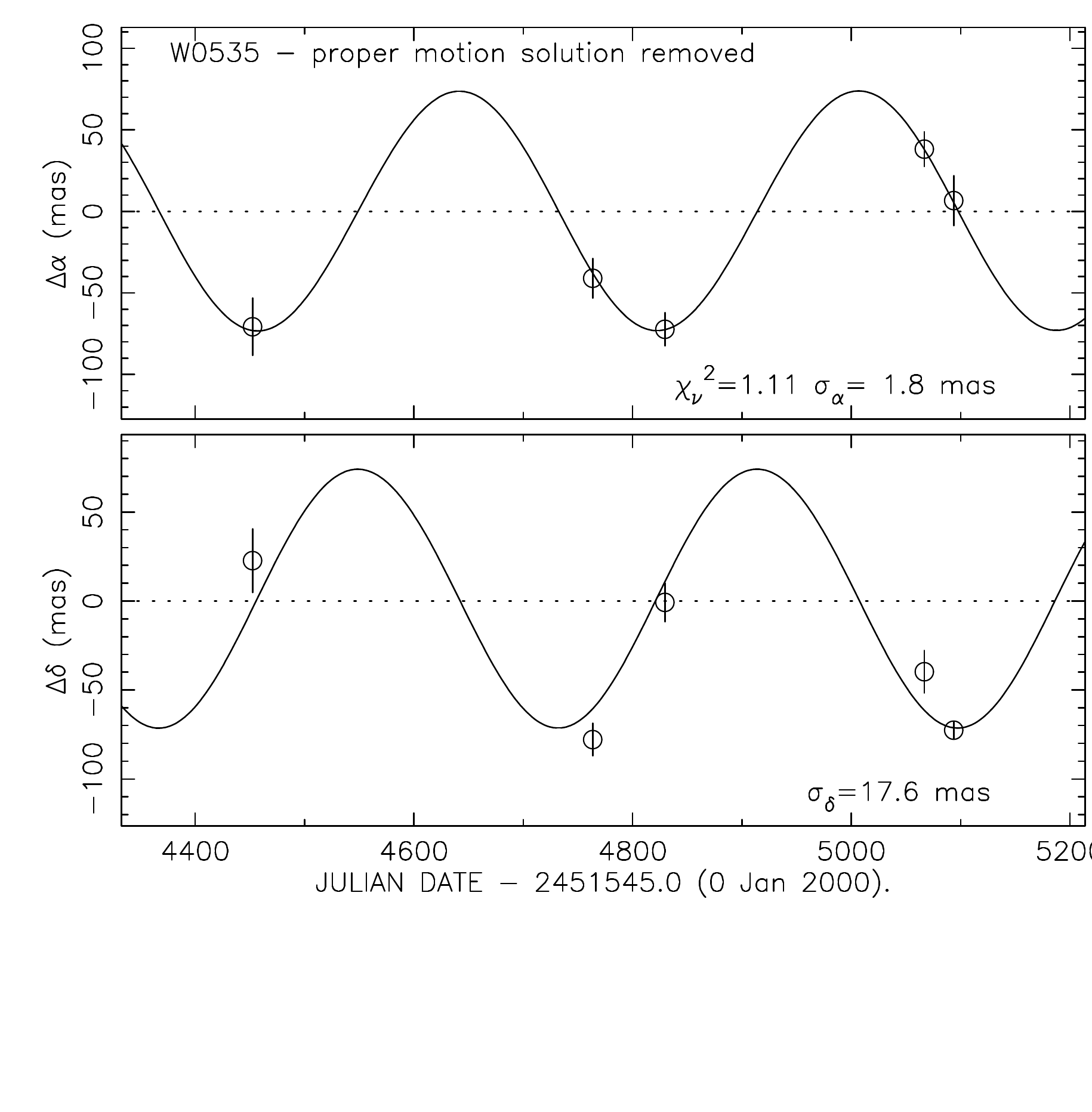}
      \includegraphics[trim=5mm 75mm 25mm 15mm, clip=true,width=65mm]{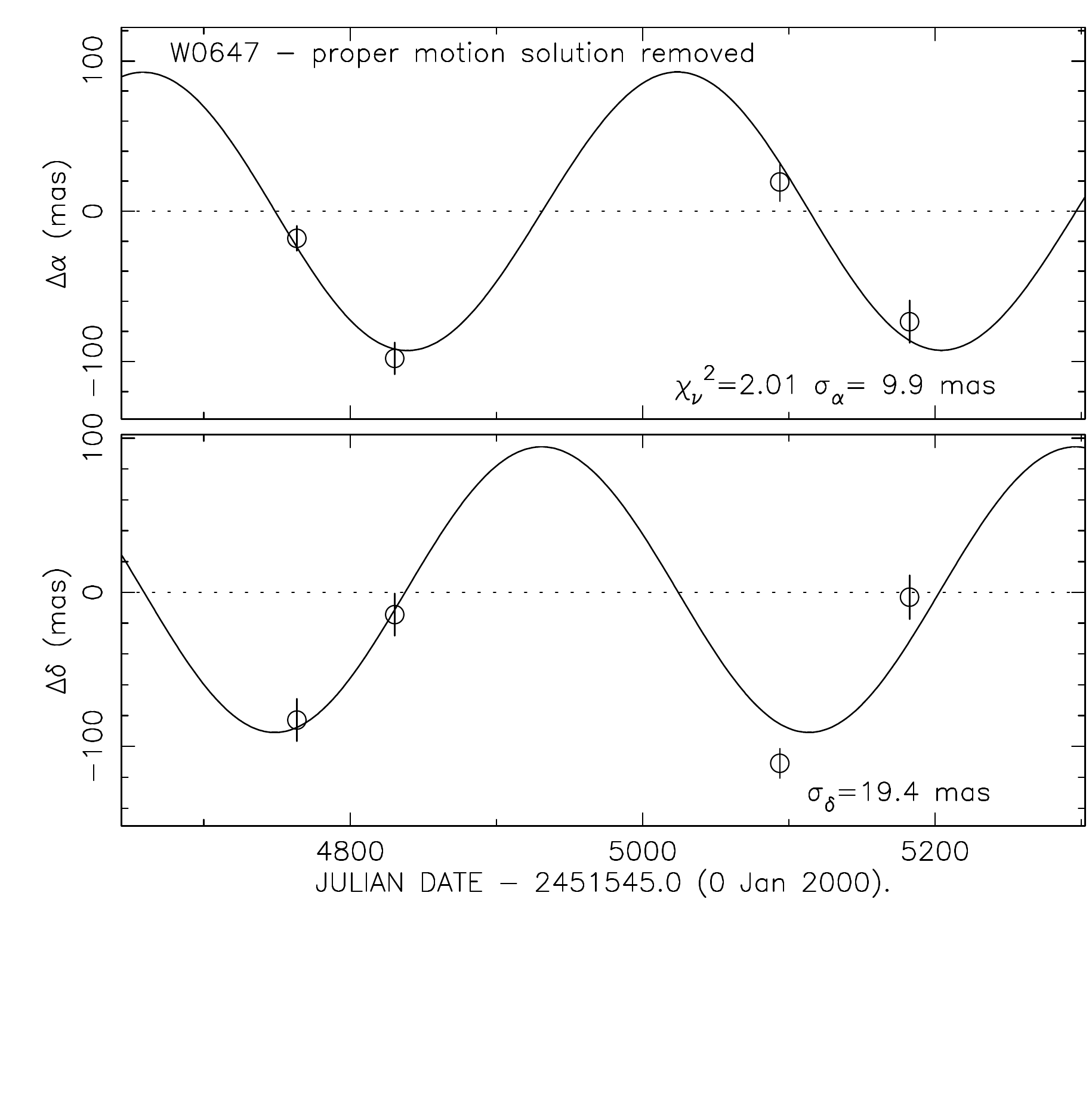}\\[-35pt]
      \includegraphics[trim=5mm 75mm 25mm 15mm, clip=true,width=65mm]{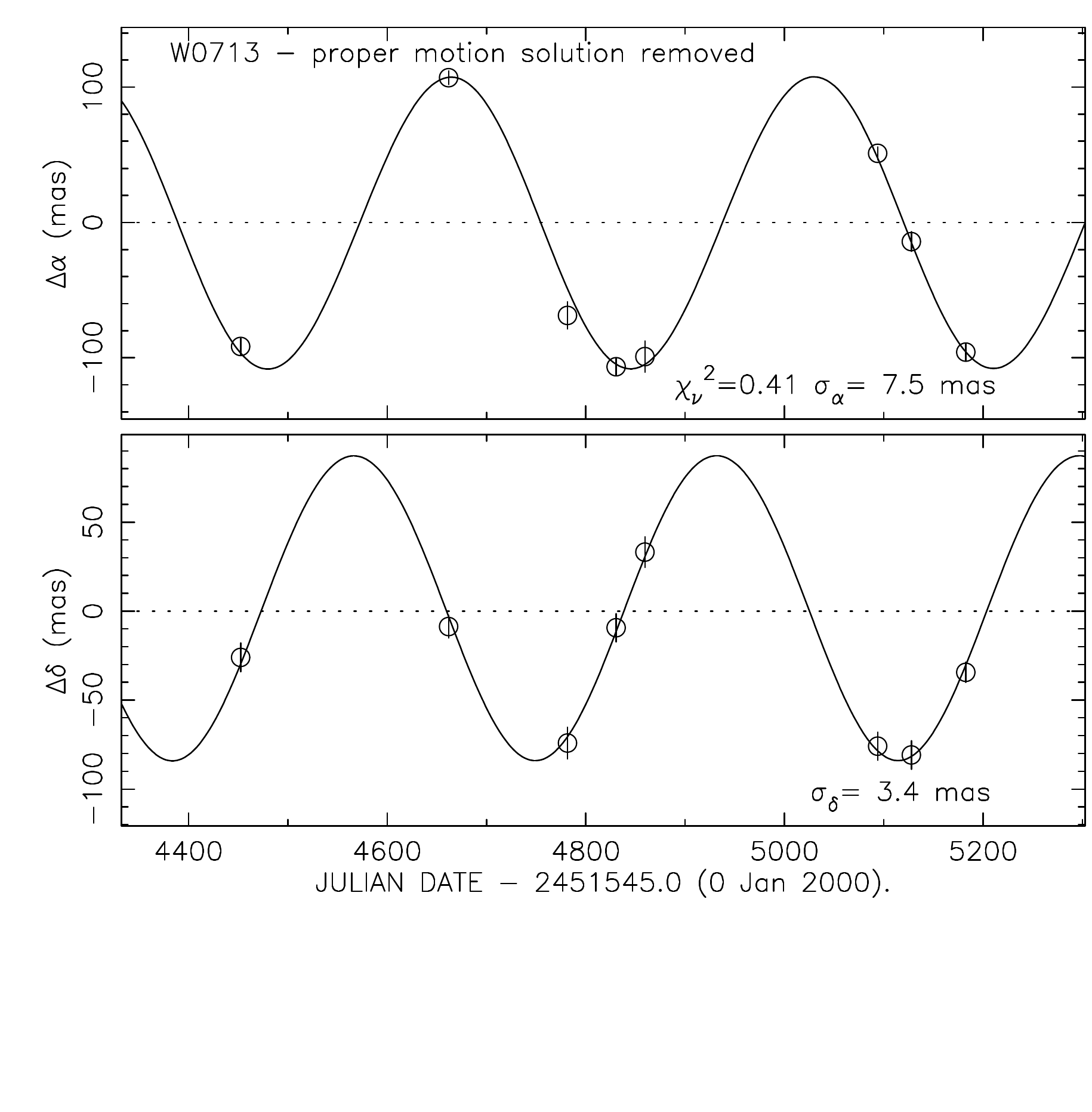}
      \includegraphics[trim=5mm 75mm 25mm 15mm, clip=true,width=65mm]{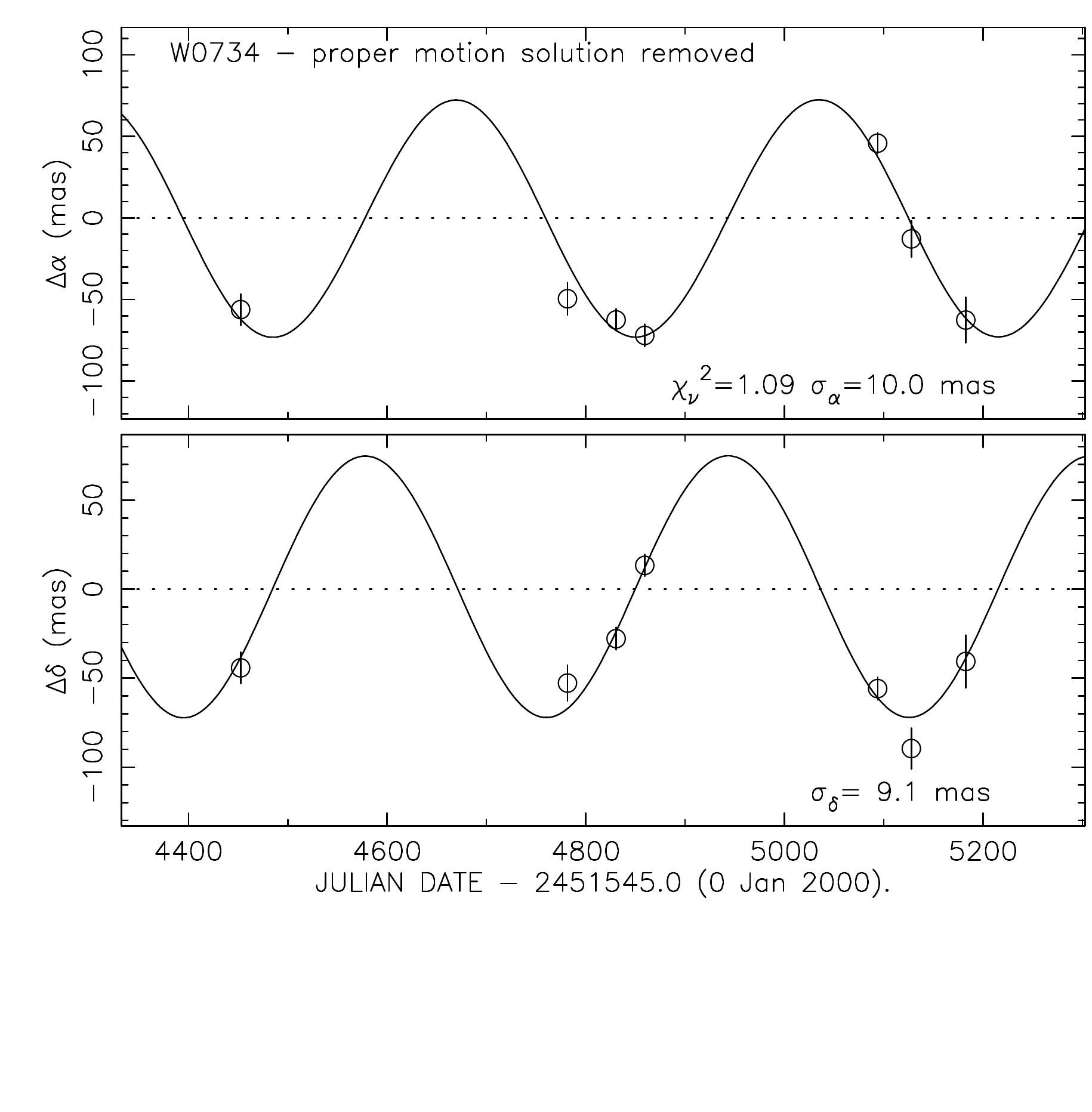}\\[-35pt]
      \includegraphics[trim=5mm 75mm 25mm 15mm, clip=true,width=65mm]{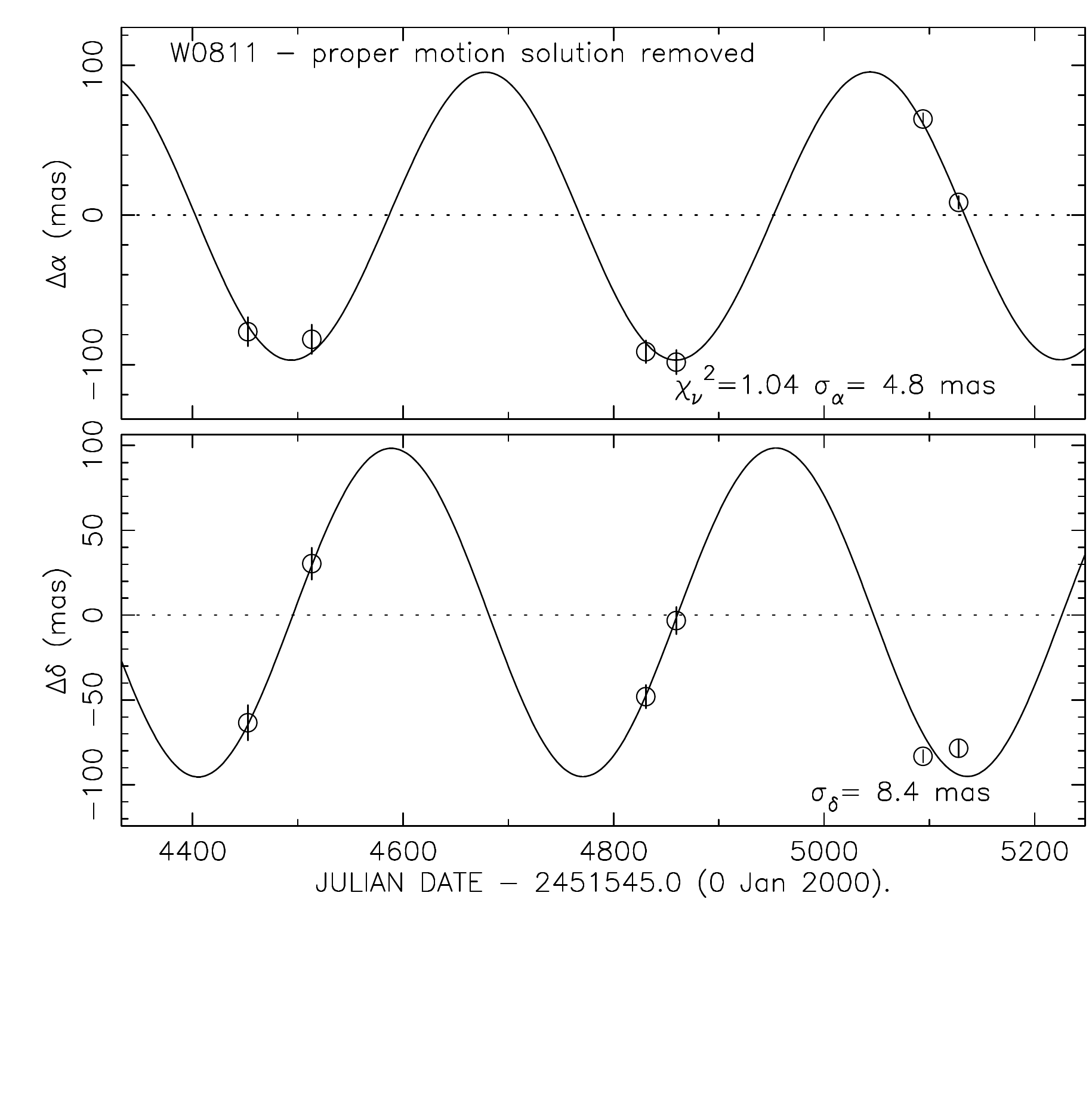}
      \includegraphics[trim=5mm 75mm 25mm 15mm, clip=true,width=65mm]{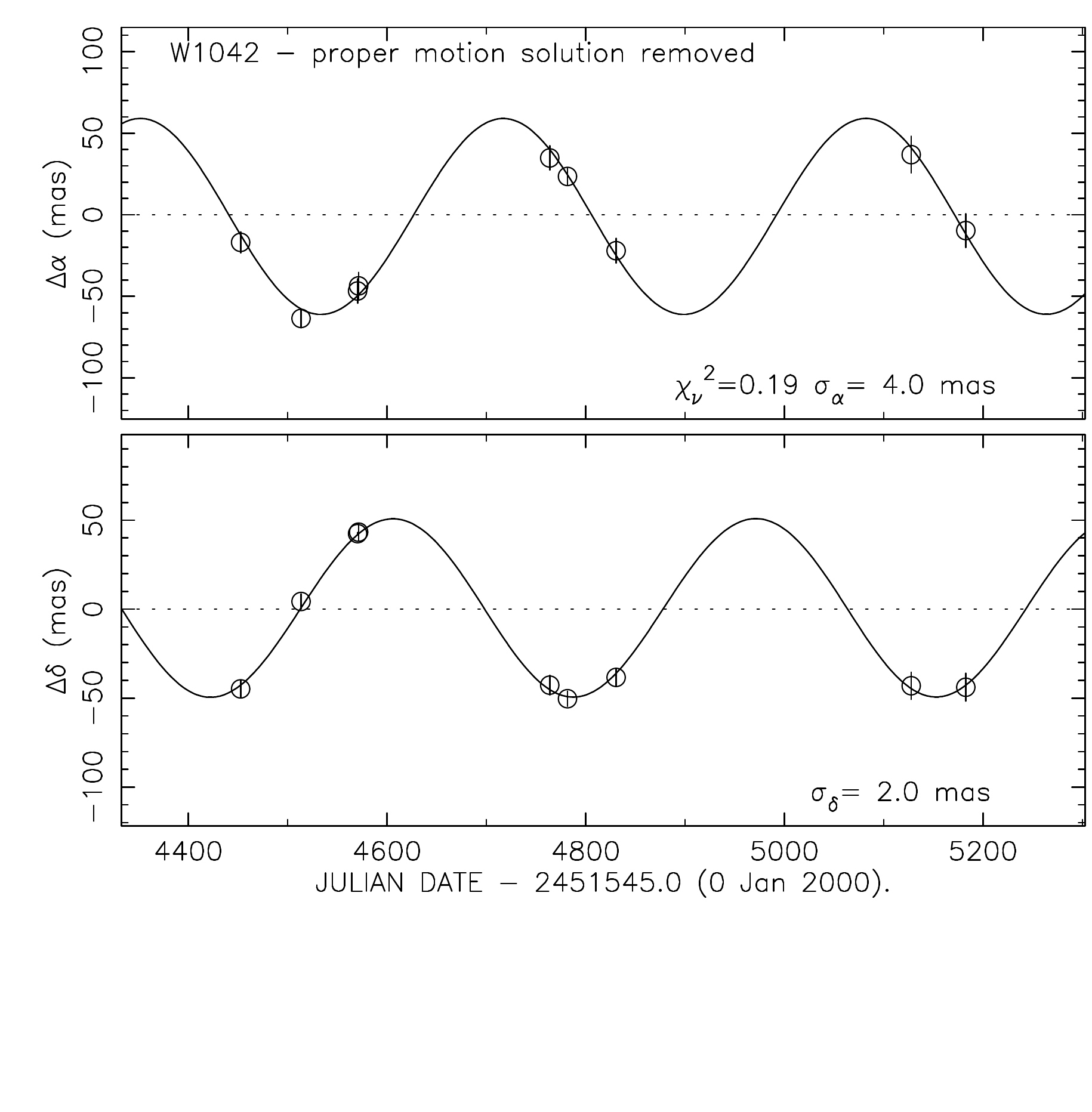}
   \end{center}
   \vspace{-1.0cm}
   \caption{Astrometric solutions as reported in Table \ref{Results}, with the fitted proper motion removed for clarity.\label{Fig0a}}
\end{figure*}

Astrometric parameters are presented in Table \ref{Results} along with the number of epochs 
(N$_\mathrm{epoch}$) included in each solution. Plots of the
astrometric solutions are shown in Fig. \ref{Fig0a} (with proper motions removed for clarity). These
show the baseline of the observations obtained, the observation
epochs for each target, and the scatter ($\sigma_\alpha, \sigma_\delta$) and reduced chi-squared ($\chi_{\nu}^2$) 
about each solution.

As W1141 has no extant spectral type, we have used its absolute magnitudes (\MWT=14.68 and \MJ=19.81) to estimate that it would
have an equivalent spectral type of Y0 $\pm$1 sub-type (the eleven other Y0 dwarfs with extant absolute magnitude
estimates have median values of \MWT=14.65 and \MJ=20.32).

The parallax estimates for W0535, W0647 and W2325 are considered preliminary, being based on less than 6 epochs and
delivering parallax precisions of greater than 10\,mas. Observations for these targets are continuing.

The quality of the solutions  is visibly quite good. There is a noticeable trend for the brightest sources 
(e.g. W0148, W0713, W1042, W1141,W2102,W2332 with J3 $\lesssim$ 19.5) to
have $\chi_{\nu}^2$ systematically below 1, suggesting that the procedure for estimating per-epoch
uncertainties for the brightest targets may be over estimating these uncertainties. The impact
of this over-estimation is that our final parallax and proper-motion uncertainties are likely to
be conservative. 

There are two solutions which require further discussion. W1639's uncertainties are 
systematically {\em underestimated} as this Y dwarf's astrometry requires a simultaneous  
point-spread function fit with a nearby much brighter star \citep[for images of the field see][]{Tinney12}. The increased uncertainties associated with this
joint solution at each epoch will not be reflected in the scatter about the (isolated) reference stars. 
Since this scatter is used to derive the per-epoch uncertainty for W1639, this will
result in underestimated per-epoch uncertainties and  an artificially high $\chi_{\nu}^2$. Similarly, W1541,  is
currently undergoing a passage past a star 2.5 magnitudes brighter and 1$\arcsec$ to the south. In this case the ``contamination''
of the astrometry is so pronounced in the declination direction that the parallax solution is determined by the
right ascension component alone -- for all other objects it is determined from both the declination and right ascension
solutions.

\begin{figure*} 
   \figurenum{2 (cont.)}
   \begin{center}
      \includegraphics[trim=5mm 75mm 25mm 15mm, clip=true,width=65mm]{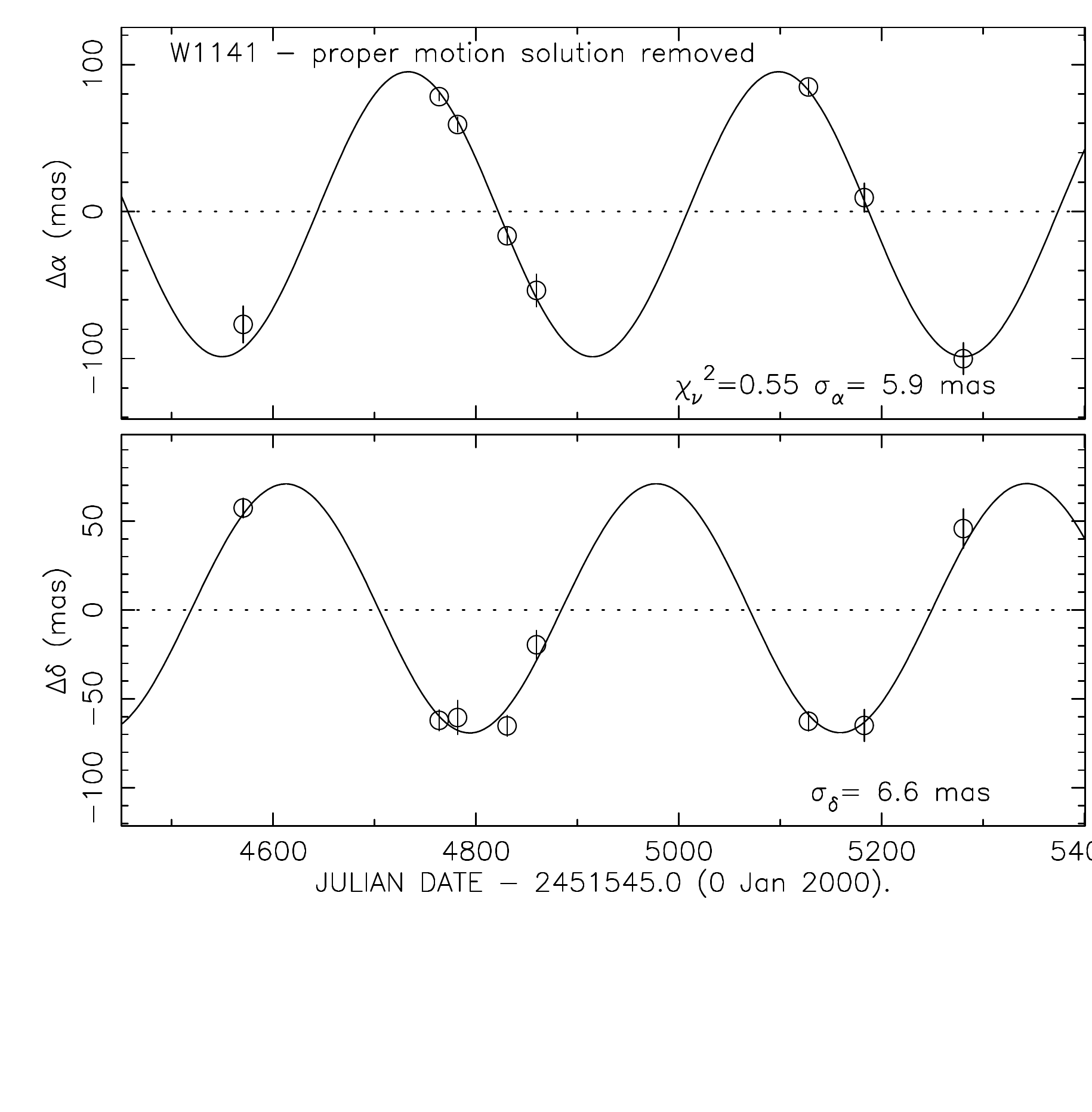}
      \includegraphics[trim=5mm 75mm 25mm 15mm, clip=true,width=65mm]{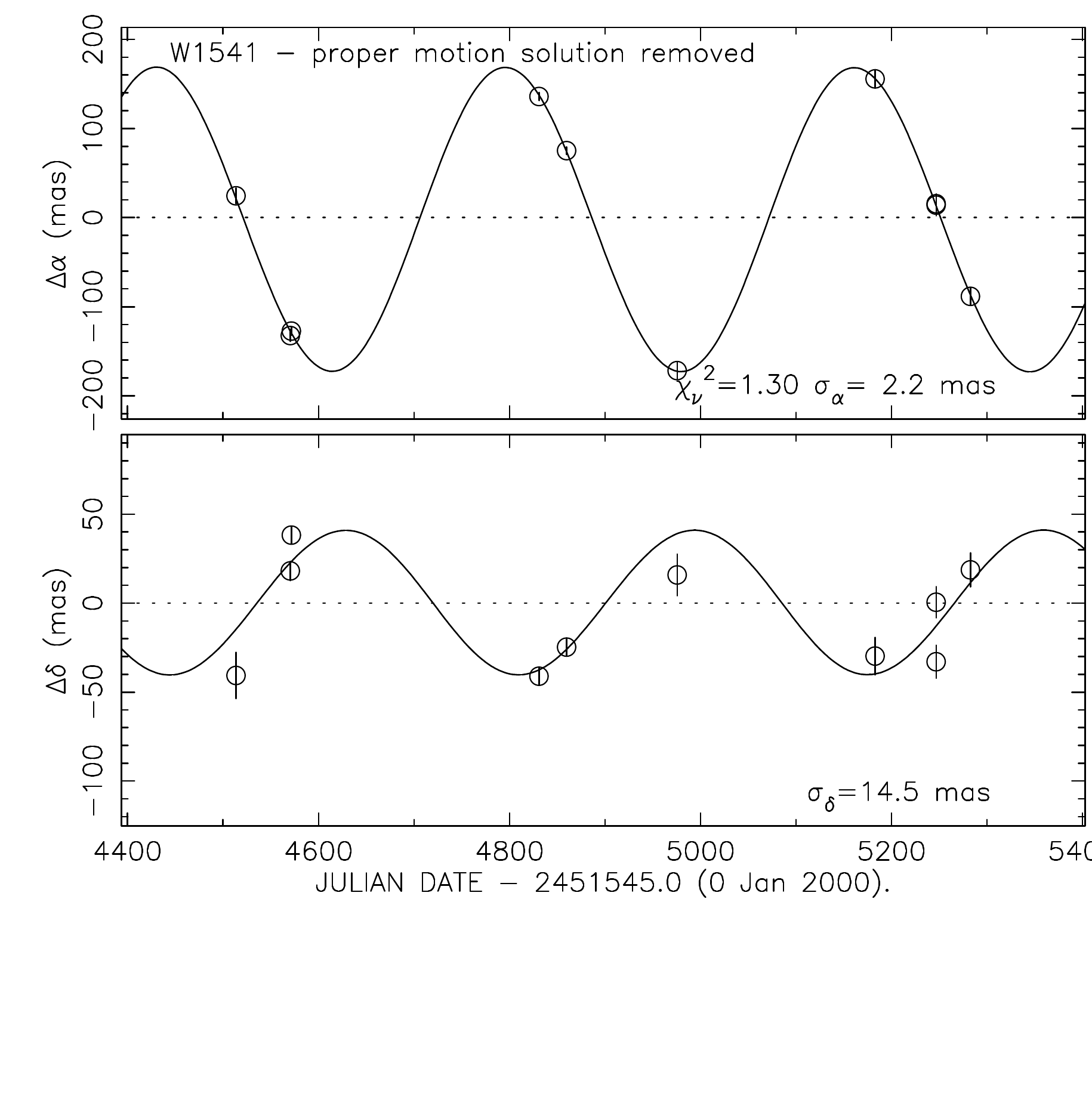}\\[-35pt]
      \includegraphics[trim=5mm 75mm 25mm 15mm, clip=true,width=65mm]{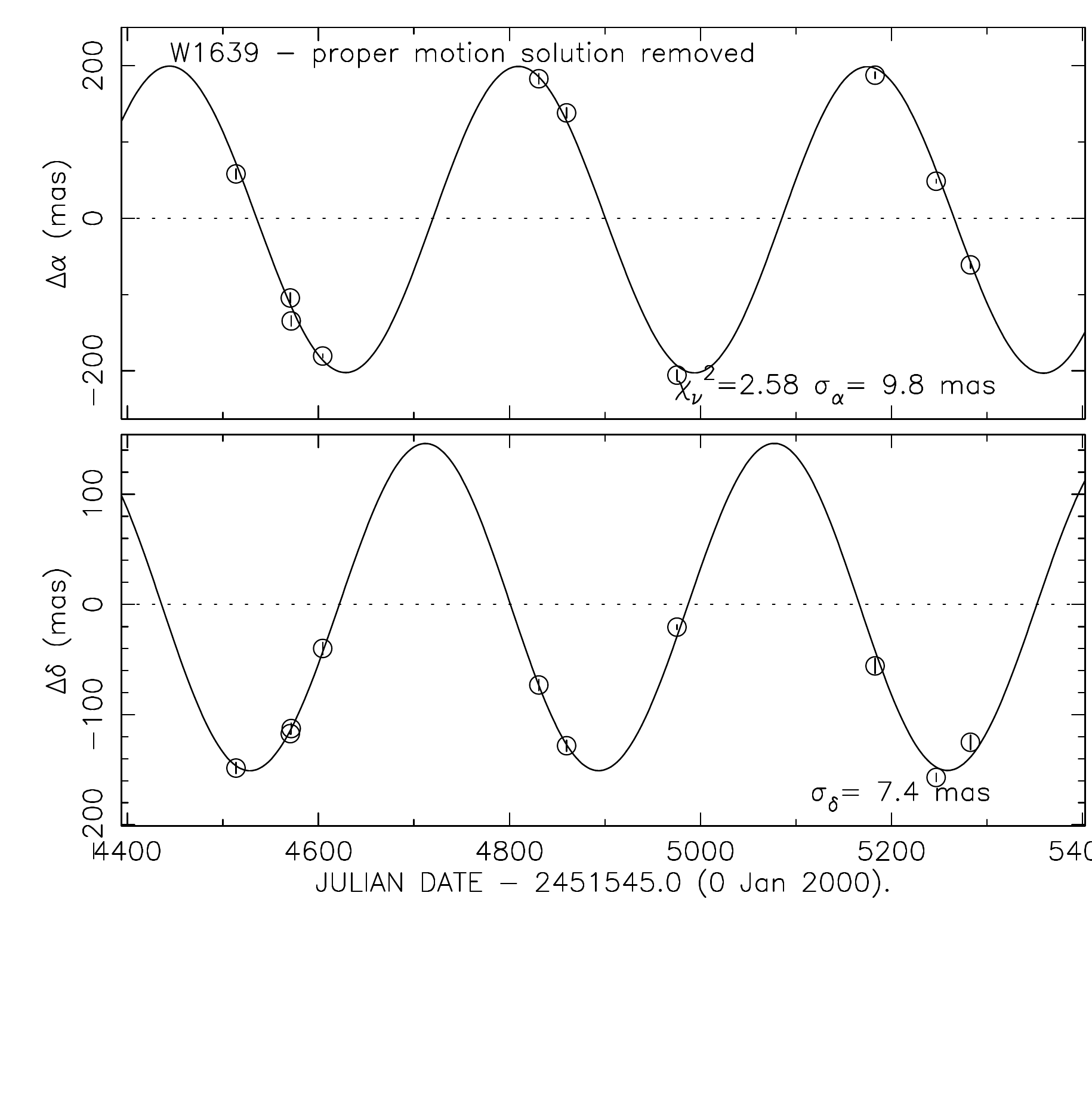}
      \includegraphics[trim=5mm 75mm 25mm 15mm, clip=true,width=65mm]{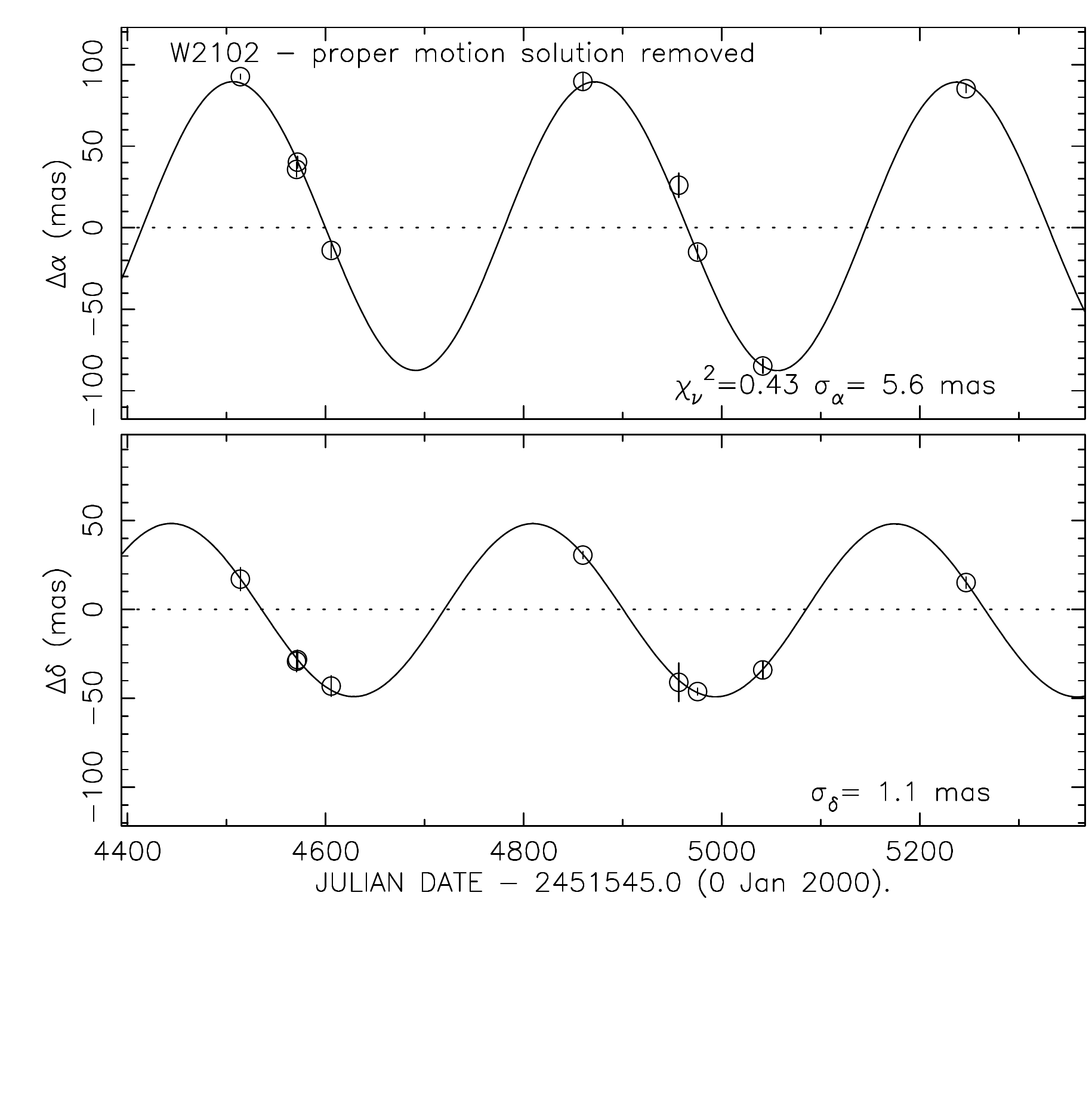}\\[-35pt]
      \includegraphics[trim=5mm 75mm 25mm 15mm, clip=true,width=65mm]{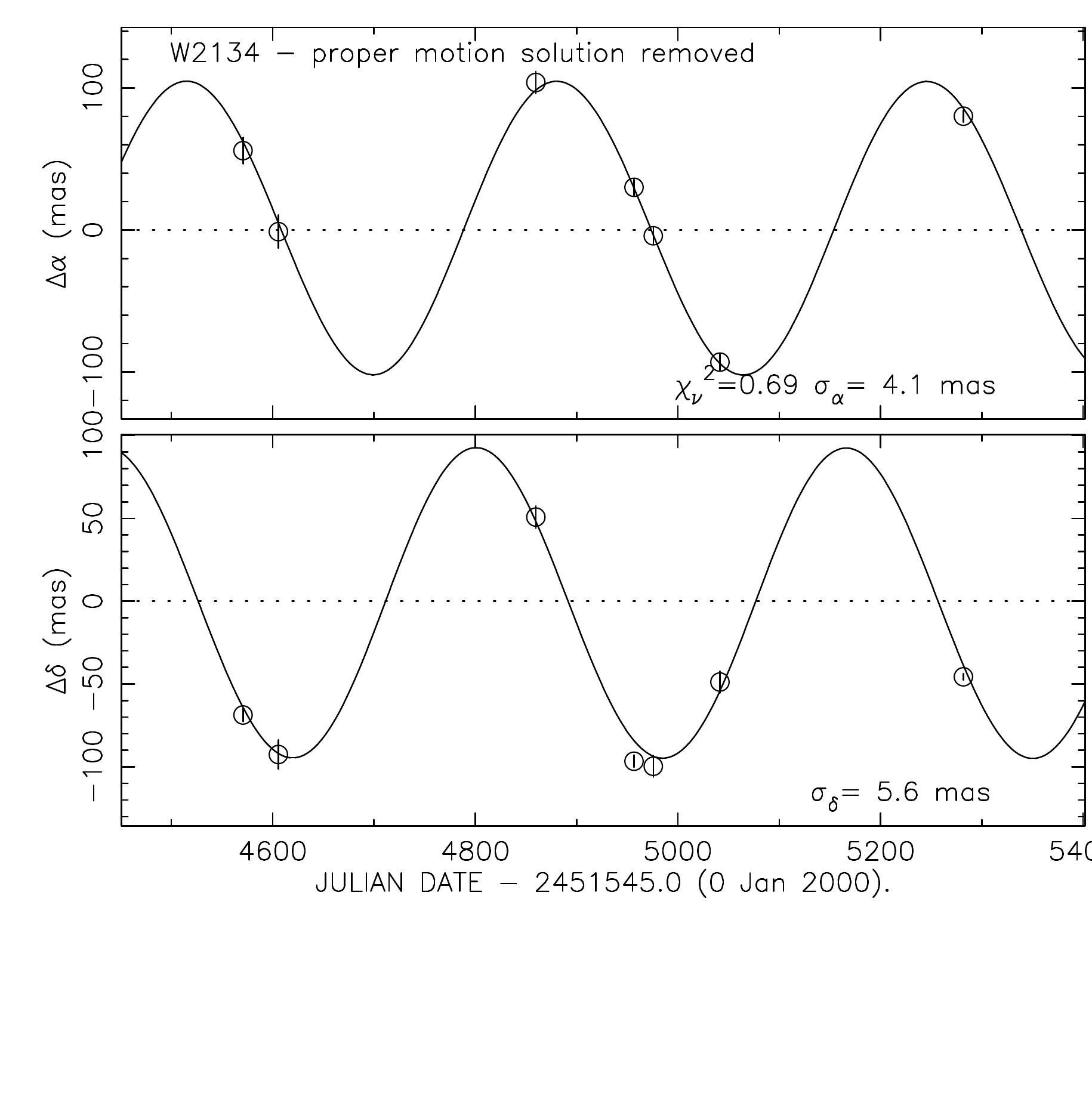}
      \includegraphics[trim=5mm 75mm 25mm 15mm, clip=true,width=65mm]{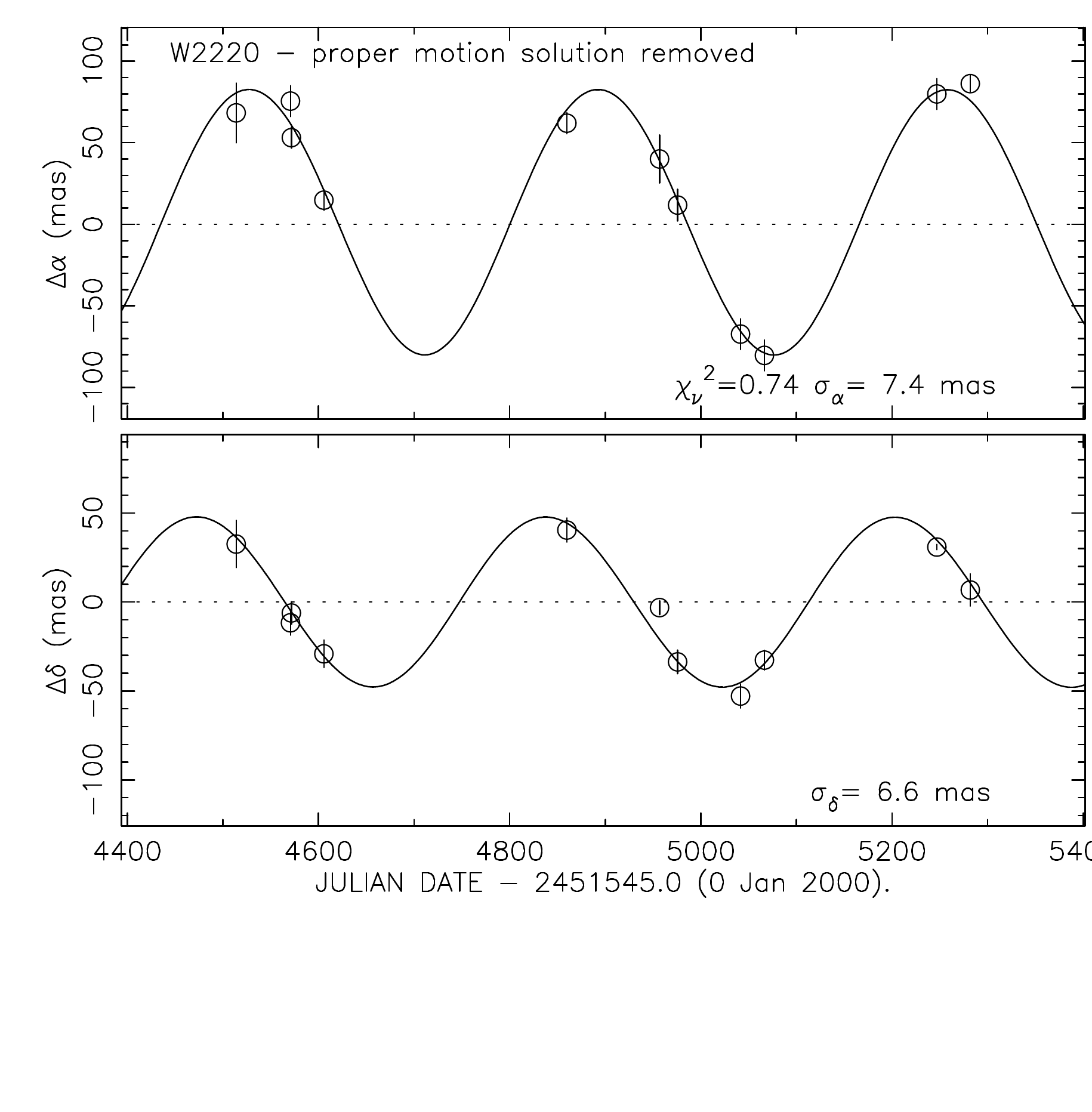}\\[-35pt]
      \includegraphics[trim=5mm 75mm 25mm 15mm, clip=true,width=65mm]{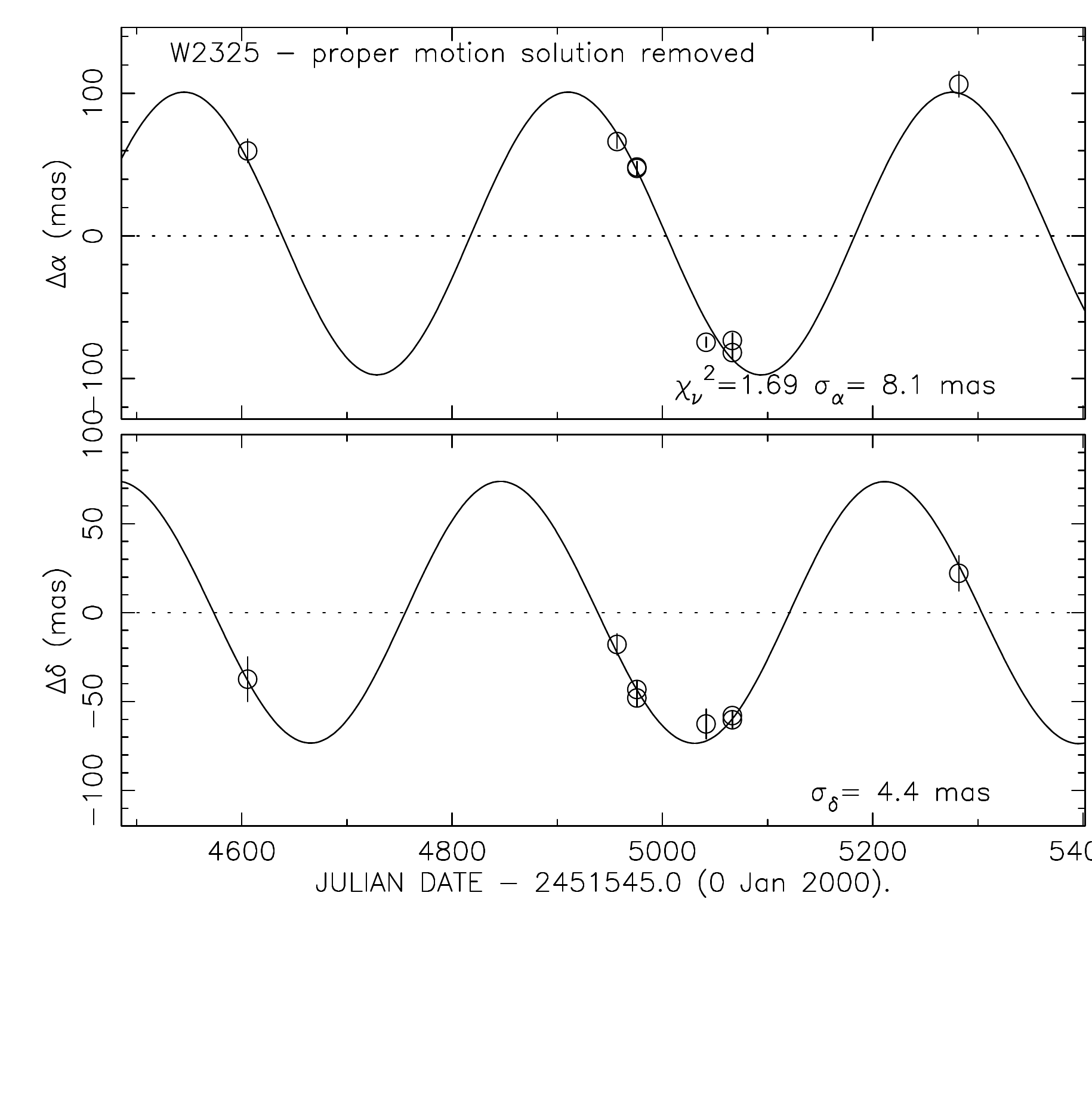}
      \includegraphics[trim=5mm 75mm 25mm 15mm, clip=true,width=65mm]{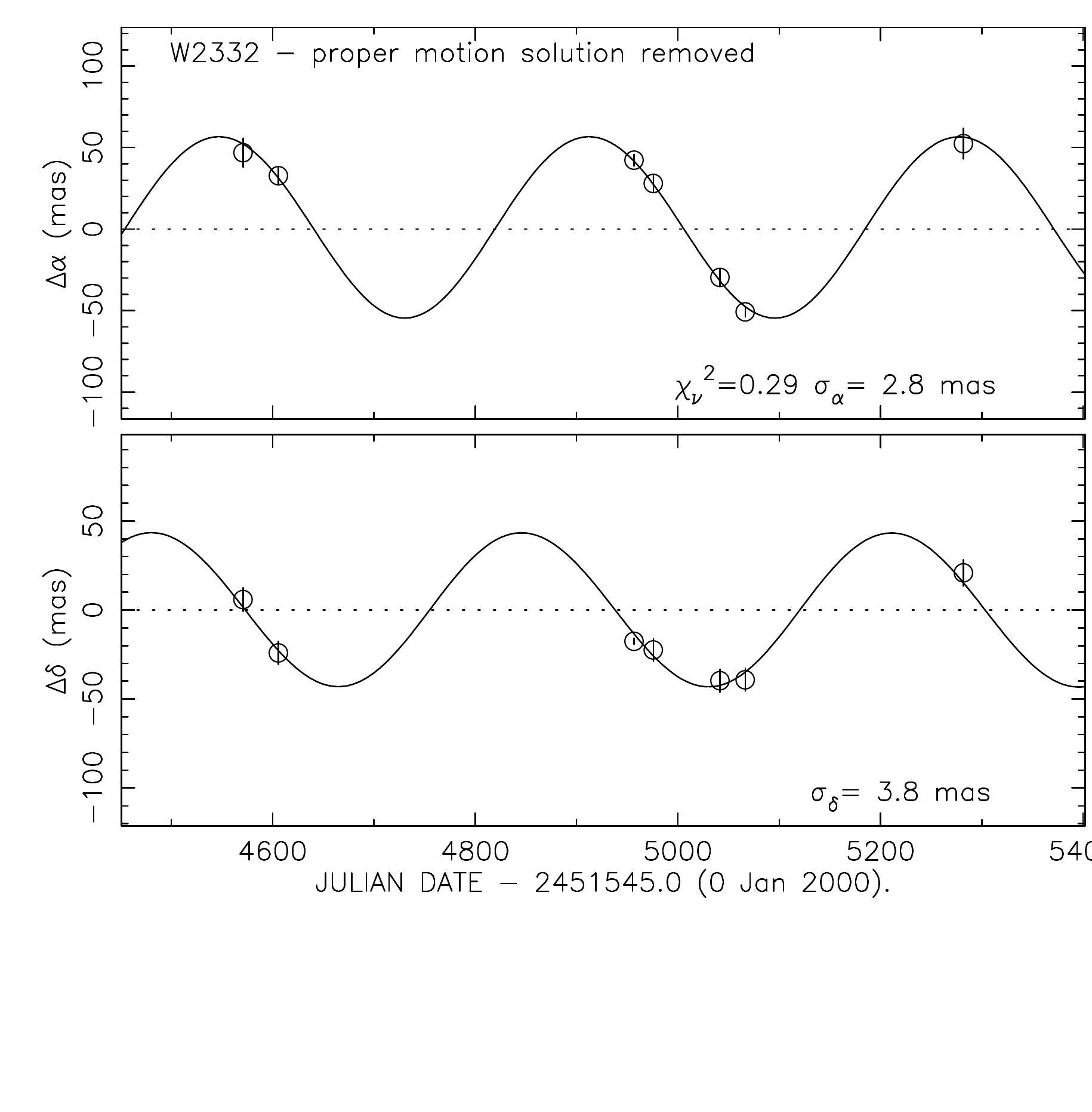}\\[-35pt]
   \end{center}
   \vspace{-1.0cm}
   \caption{Astrometric solutions as reported in Table \ref{Results}, with the fitted proper motion removed for clarity.\label{Fig0b}}
\end{figure*}

\subsection{Other Data\label{other text}}

Table \ref{Otherdata} provides photometric and parallax data for Y dwarfs with previously published parallaxes due to \citet[hereafter D\&K13][]{DK13}, 
\citet[hereafter K13] {Kirkpatrick13a}, \citet[hereafter B14]{Beichman14} and most recently
for W0855 by \citet[hereafter L14]{Luhman14}, as well as for a reference sample of previously published late-T dwarfs. 
The reference sample of late T dwarfs has been extracted from the literature focussing on those objects with extant spectra 
(and so spectral types), parallaxes with better than 10\% precision and J$_{MKO}$ photometry. 
It does not aim to be a complete sample of all T dwarfs, but rather a comparison sample of known T dwarfs with high quality data.
Sources for the astrometry, spectral types and photometry for these objects are provided in the Table.
Photometry for these objects tends to fall into two classes: lower precision data acquired to identify and classify sources
via J--W2 colours \citep[e.g.][]{kirkpatrick2012,Cushing11} and deeper data acquired to obtain
better photometry \citep[e.g.][]{Leggett13}. Whenever the latter is available we have preferred it in the Table.  
For w0855 we use the 2.6\,$\sigma$ J3 detection of \citet{Faherty14b} to estimate a J\mko\ magnitude, and
adopt the uncontaminated W2 magnitude and updated parallax from \citet{Wright14}.
In the absence of any spectrum for W0855, we assign it a spectral type of $>$Y2 (similar to that
assigned to W1828) based on its very faint \MWT\ and extremely red J--W2 colour (J--W2 = 11.1$\pm$.5).  

We now consider the overlap between our new data and comparable, extant Y dwarf parallaxes summarised in Table \ref{Comp}. 
We  do not, however, compare with the results in \cite{Marsh13} -- while that astrometry was the best available at the time of publication,
it is not comparable in precision to that being examined here.

{\em B14:} We have three objects in common with B14: W0713 for which we measure $\pi$=108.7$\pm$4.0\,mas compared with B14's
106$\pm$13\,mas; W1541 for which we measure $\pi$=175.1$\pm$4.4\,mas compared with B14's 176$\pm$9\,mas; and W2220 for which we
measure $\pi$=87.2$\pm$3.7\,mas compared with B14's 106$\pm$24\,mas. In each case our solution is formally consistent (at the
combined 1-$\sigma$ level) with that of B14. Our quoted uncertainties are smaller, which is to be expected given the
per-epoch precisions obtained from the {\em Spitzer} data that dominate the B14 solutions, where the average precision
per epoch of 60\,mas compares with the median precision for our targets of 9.9\,mas. 

{\em K13}: have presented a parallax for W0647 of 115$\pm$12\,mas, while we measure $\pi$=93$\pm$13\,mas, making the two solutions
consistent at the combined 1-$\sigma$ level. Our parallax is a preliminary one based on 4 epochs (though those epochs are
well placed to measure a robust parallax). The addition of
extra epochs over the coming 12 months  will see this result become significantly better than is possible with {\em Spitzer}-based
data.
    
{\em D\&K13}: We have two objects in common with D\&K13: W1541 for which we measure $\pi$=175.1$\pm$4.4\,mas compared with D\&K13's
74$\pm$31\,mas; and W0148 for which we measure $\pi$=91.1$\pm$2.4\,mas compared with D\&K's 60$\pm$16\,mas. The difference for W1541 is
formally a greater than 3-$\sigma$ one, which we ascribe to the degeneracy present in the D\&K13 solution because it is based
on only 3 well-separated epochs (see their Fig. S1), with the earliest and poorest epoch primarily determining the
proper motion in their solution, while the later epochs constrain the parallax. The result is a solution with a
large relative uncertainty, and significant degeneracy between the proper motion and parallax parameters. Our own
data is completely inconsistent with a parallax as small as 70\,mas, and is in agreement with the larger parallax
reported by B14. We therefore do not plot the D\&K13 result for W1541 or report it in the Tables in this paper.

The situation for W0148 (where the difference is formally a 1.9\,$\sigma$ one) shares some similarities with that for W1541
in that D\&K13 report only 4 widely separated epochs and again the earliest (and by far the poorest epoch) must
strongly determine the proper motion solution, generating the possibility of degeneracy between the proper motion
and parallax. However, we do not consider the 1.6\,$\sigma$ difference between the solutions to be significant
enough to warrant rejecting either, and so report and plot both values in the Tables and Figures.

{\em W1639}: A preliminary parallax of 200$\pm$20\,mas for W1639 was previously reported by us \citep{Tinney12}. The uncertainty in that
measurement was artificially inflated to account for the added systematic uncertainties introduced by the need to
combine early FourStar data with astrometry from the WISE satellite itself. Our new measurement of 202.3$\pm$3.1\,mas 
(based on FourStar data only) confirms the previous measurement, and improves the precision significantly.

\begin{deluxetable*}{lllrrrccccc}
\tabletypesize{\scriptsize}
\tablecaption{Photometric and Astrometric data on T and Y dwarfs with published parallaxes\tablenotemark{a}\label{Otherdata}}
\tablenum{3}
\tablewidth{0pt}
\tablehead{
\colhead{Shorthand}   & \colhead{\em AllWISE} &\colhead{SpT}  &\colhead{$\pi$}&\colhead{J$_{MKO}$}   &\colhead{W2}   &\colhead{SpT}&\colhead{$\pi$}&\colhead{Phot} &\colhead{Disc} \\
\colhead{Designation} & \colhead{Designation} &               &\colhead{(mas)}&\colhead{(mag)}       &\colhead{(mag)}&\colhead{Ref}&\colhead{Ref}  &\colhead{Ref}  &\colhead{Ref}
}
\startdata
	    &  {\bf Y and Late-T dwarfs} &      &               &                 &                  &     &   &    &     \\
 W0146 &  WISEA J014656.66+423409.9 & Y0   &  94$\pm$14    & 19.40$\pm$0.25  & 15.083$\pm$0.065 &  1  &11 &1   &  4  \\
 W0148 &  WISEA J014807.34-720258.7 & T9.5 &  60$\pm$16    & 18.96$\pm$0.07  & 14.592$\pm$0.039 &  2  &12 &2   &  4  \\
 W0254 &  WISEA J025409.55+022358.5 & T8   &  135$\pm$15   & 16.14$\pm$0.12  & 12.758$\pm$0.026 &  2  &12 &3   &  2,37  \\
 W0313 &  WISEA J031326.00+780744.3 & T8.5 & 153$\pm$15    & 17.67$\pm$0.07  & 13.263$\pm$0.026 &  2  &11 &1   &  2  \\
 W0335 &  WISEA J033515.07+431044.7 & T9   &  70$\pm$9     & 20.07$\pm$0.30  & 14.515$\pm$0.055 &  4  &11 &4   &  4  \\
 W0410 &  WISEA J041022.75+150247.9 & Y0   &  160$\pm$09   & 19.44$\pm$0.03  & 14.113$\pm$0.047 &  5  &11 &21  &  2,5  \\
       &                            &      &  132$\pm$15   &                 &                  &     &12 &    &     \\
 W0647 &  WISEA J064723.24-623235.4 & Y1   &  115$\pm$12   & 22.65$\pm$0.27  & 15.224$\pm$0.051 &  32 &32 &32  &  32 \\
 W0713 &  WISEA J071322.55-291752.0 & Y0   &  106$\pm$13   & 19.64$\pm$0.15  & 14.462$\pm$0.052 &  1  &11 &1   &  1  \\
 W1311 &  WISEA J131106.21+012253.9 & T9   &  62$\pm$12    & 18.75$\pm$0.07  & 14.703$\pm$0.060 &  2  &11 &2   &  2  \\
 W0855 &  WISE J085510.83-071442.5  & ($>$Y2)\tablenotemark{b}
                                    &  448$\pm$33  & 25.00$^{+0.53}_{-0.35}$ & 14.02$\pm$0.05   &     &47 &47  &  35 \\
 W1405 &  WISEA J140518.32+553421.3 & Y0p  &  129$\pm$19   & 21.06$\pm$0.06  & 14.097$\pm$0.037 &  5  &12 &21  &  2,5  \\
 W1541 &  WISEP J154151.65-225025.2 & Y0.5 &  176$\pm$9    & 21.12$\pm$0.06  & 13.982$\pm$0.112 &  5  &11 &21  &  2,5   \\
 W1542 &  WISEA J154214.00+223005.2 & T9.5 &  96$\pm$41    & 20.25$\pm$0.13  & 15.043$\pm$0.061 &  4  &11 &4   &  4   \\
 W1738 &  WISEA J173835.52+273258.8 & Y0   &  128$\pm$10   & 20.05$\pm$0.09  & 14.497$\pm$0.043 &  5  &11 &21  &  2,5  \\
       &                            &      &  102$\pm$18   &                 &                  &     &12 &    &     \\
 W1741 &  WISEA J174124.22+255319.2 & T9   &  180$\pm$15   & 16.18$\pm$0.02  & 12.347$\pm$0.023 &  2  &12 &2   &  2,37   \\
 W1804 &  WISEA J180435.37+311706.2 & T9.5:&  80$\pm$10    & 18.67$\pm$0.04  & 14.590$\pm$0.046 &  2  &11 &2   &  2   \\
       &                            &      &  60$\pm$11    &                 &                  &     &12 &    &     \\
 W1828 &  WISEA J182831.08+265037.6 & $>$Y2&  106$\pm$7    & 23.48$\pm$0.23  & 14.353$\pm$0.045 &  1  &11 &21  &  2,5   \\
       &                            &      &  70$\pm$14    &                 &                  &     &12 &    &     \\
 W2056 &  WISEA J205628.88+145953.6 & Y0   &  140$\pm$09   & 19.43$\pm$0.04  & 13.839$\pm$0.037 &  5  &11 &21  &  2,5   \\
       &                            &      &  144$\pm$23   &                 &                  &     &12 &    &     \\
 W2209 &  WISEA J220905.75+271143.6 & Y0:  &  147$\pm$11   & 22.58$\pm$0.14  & 14.770$\pm$0.055 &  33 &11 &33  &  2   \\
 W2220 &  WISEA J222055.34-362817.5 & Y0   &  106$\pm$24   & 20.38$\pm$0.17  & 14.714$\pm$0.056 &  1  &11 &1   &  1  \\[6pt]
       &  {\bf Reference T dwarfs } &      &               &                 &                  &     &   &    &     \\
 U0034 &  ULAS J003402.77-005206.7  & T8.5 &  68.7$\pm$1.4 & 18.15$\pm$0.03  & 14.500$\pm$0.079 &  6,5&13 &6   &  6   \\
2M0034 & 2MASS J00345157+0523050    & T6.5 &  105$\pm$8    & 15.11$\pm$0.03  & 12.520$\pm$0.028 &  7  &14 &10  &  39   \\
2M0050 & 2MASS J00501994-3322402    & T7   &  94.6$\pm$2.4 & 15.65$\pm$0.10  & 13.550$\pm$0.036 &  7  &13 &15  &  38   \\
CF0059 & CFBDS J005910.90-011401.3  & T8.5 & 103.2$\pm$2.1 & 18.06$\pm$0.03  & 13.681$\pm$0.043 &  5  &13 &23  &  23   \\
2M0243 &    2MASSI J0243137-245329  & T6   &  94$\pm$4     & 15.13$\pm$0.03  & 12.923$\pm$0.027 &  7  &15 &24  &  40   \\
2M0415 &    2MASSIJ0415195-093506   & T8   & 175.2$\pm$1.7 & 15.32$\pm$0.03  & 12.261$\pm$0.026 &  7  &13 &24  &  40   \\
2M0559 &    2MASSI J0559191-140448  & T4.5 &  96.6$\pm$1.0 & 13.57$\pm$0.03  & 11.904$\pm$0.023 &  7  &13 &28  &  41   \\
U0722	 & UGPS J072227.51-054031.2   & T9   & 242.8$\pm$2.4 & 16.52$\pm$0.02  & 12.213$\pm$0.027 &  5  &16 &25  &  25  \\
2M0727 &    2MASSI J0727182+171001  & T7   & 112.5$\pm$0.9 & 15.19$\pm$0.03  & 12.962$\pm$0.033 &  7  &13 &24  &  40   \\
2M0729 &    2MASS J07290002-3954043 & T8p  &  126$\pm$8    & 15.64$\pm$0.08  & 12.964$\pm$0.026 &  8  &14 &26  &  8   \\
U0901	 & ULAS J090116.23-030635.0   & T7.5 &  62.6$\pm$2.6 & 17.90$\pm$0.04  & 14.604$\pm$0.074 &  9  &17 &9   &  9   \\
2M0939 &    2MASS J09393548-2448279 & T8   &  187$\pm$5    & 15.61$\pm$0.09  & 11.639$\pm$0.022 &  7  &18 &27  &  38  \\
2M1007 &    2MASS J10073369-4555147 & T5   &  71$\pm$5     & 15.42$\pm$0.07  & 13.870$\pm$0.040 &  8  &14 &26  &  8   \\
2M1047 &    2MASSI J1047538+212423  & T6.5 &  95$\pm$4     & 15.46$\pm$0.03  & 12.972$\pm$0.032 &  7  &15 &28  &  42   \\
S1110	 & SDSS J111010.01+011613.1   & T5.5 &  52.1$\pm$1.2 & 16.12$\pm$0.05  & 13.917$\pm$0.047 &  7  &13 &28  &  43   \\
2M1114 &    2MASS J11145133-2618235 & T7.5 & 179.2$\pm$1.4 & 15.52$\pm$0.05  & 12.239$\pm$0.026 &  7  &13 &22  &  38   \\
2M1217 &    2MASSI J1217110-031113  & T7.5 &  90.8$\pm$2.2 & 15.56$\pm$0.03  & 13.197$\pm$0.035 &  7  &19 &28  &  42   \\
2M1237 &    2MASS J12373919+6526148 & T6.5 &  96$\pm$5     & 15.56$\pm$0.10  & 12.946$\pm$0.027 &  7  &15 &22  &  42   \\
U1335	 & ULAS J133553.45+113005.2   & T8.5 &  99.9$\pm$1.6 & 17.90$\pm$0.01  & 13.865$\pm$0.042 &  5  &13 &29  &  29   \\
S1346	 & SDSSp J134646.45-003150.4  & T6.5 &  68.3$\pm$2.3 & 15.49$\pm$0.05  & 13.567$\pm$0.034 &  7  &19 &30  &  30  \\
Gl570D & Gl570D                     & T7.5 & 171.2$\pm$0.9 & 14.82$\pm$0.05  & 12.114$\pm$0.023 &  7  &20 &34  &  44  \\
2M1503 &    2MASSW J1503196+252519  & T5   & 157.2$\pm$2.2 & 13.55$\pm$0.03  & 11.723$\pm$0.021 &  7  &13 &24  &  45   \\
S1504	 & SDSS J150411.63+102718.3   & T7   &  46.1$\pm$1.5 & 16.49$\pm$0.03  & 14.062$\pm$0.040 &  10 &13 &10  &  10  \\
2M1546 &    2MASS J15462718-3325111 & T5.5 &  88.0$\pm$1.9 & 15.41$\pm$0.05  & 13.445$\pm$0.037 &  7  &19 &26  &  40   \\
2M1615 &    2MASS J16150413+1340079 & T6   &  69$\pm$6     & 16.11$\pm$0.09  & 14.194$\pm$0.053 &  8  &14 &26  &  8  \\
S1624	 & SDSSp J162414.37+002915.6  & T6   &  90.9$\pm$1.2 & 15.20$\pm$0.05  & 13.085$\pm$0.032 &  7  &19 &31  &  31   \\
S1628	 & SDSS J162838.77+230821.1   & T7   &  75.1$\pm$0.9 & 16.25$\pm$0.03  & 13.961$\pm$0.043 &  10 &13 &10  &  10   \\
S1758	 & SDSS J175805.46+463311.9   & T6.5 &  71.0$\pm$1.9 & 15.86$\pm$0.03  & 13.823$\pm$0.032 &  7  &20 &24  &  24   \\
2M1828 &    2MASS J18283572-4849046 & T5.5 &  84$\pm$8     & 14.94$\pm$0.06  & 12.773$\pm$0.029 &  7  &14 &26  &  39   \\
2M2228 &    2MASS J22282889-4310262 & T6   &  94$\pm$7     & 15.42$\pm$0.07  & 13.328$\pm$0.035 &  7  &14 &26  &  46   \\
2M2356 &    2MASSI J2356547-155310  & T5.5 &  69$\pm$3     & 15.48$\pm$0.03  & 13.708$\pm$0.042 &  7  &15 &24  &  40   \\
\enddata                                                                                                        
\tablenotetext{a}{Spectral type (SpT), parallax ($\pi$) and photometry in the J\mko\ and {\em AllWISE} W2 passbands. No spectrum has been
published for W0855, so its type estimate ($>$Y2) is based on the lower limit to its J--W2 colour, together with its very low
luminosity in  W2 and J.}
\tablenotetext{b}{The upper limit to the spectral type for W0855 is an estimate based on its photometry and extremely faint absolute magnitude.}
\tablerefs{
  1~-~\citet{kirkpatrick2012};
  2~-~\citet{kirkpatrick2011};
  3~-~\citet{Liu11};
  4~-~\citet{Mace13};
  5~-~\citet{Cushing11};
  6~-~\citet{Warren07};
  7~-~\citet{Burgasser06};
  8~-~\citet{Looper07};
  9~-~\citet{Lodieu07};
 10~-~\citet{Chiu06};
 11~-~\citet{Beichman14};
 12~-~\citet{DK13};
 13~-~\citet{DL12};
 14~-~\citet{Faherty12};
 15~-~\citet{Vrba04};
 16~-~\citet{Leggett12};
 17~-~\citet{Marocco10};
 18~-~\citet{Burgasser08};
 19~-~\citet{Tinney03};
 20~-~\citet{van Leeuwen07};
 21~-~\citet{Leggett13};
 22~-~\citet{Leggett10};
 23~-~\citet{Delorme08};
 24~-~\citet{Knapp04};
 25~-~\citet{Lucas10};
 26~-~Synthetic photometry derived from literature spectra by Trent Dupuy (\url{https://www.cfa.harvard.edu/~tdupuy/plx/Database\_of\_Ultracool\_Parallaxes.html});
 27~-~\citet{Leggett09};
 28~-~\citet{Leggett02};
 29~-~\citet{Burningham08};
 30~-~\citet{Tsvetanov00};
 31~-~\citet{Strauss99};
 32~-~\citet{Kirkpatrick13a};
 33~-~\citet{Cushing14b};
 34~-~\citet{Geballe01};
 35~-~\citet{Luhman14};
 36~-~\citet{Faherty14b};
 37~-~\citet{Scholz11};
 38~-~\citet{Tinney05};
 39~-~\citet{Burgasser04}; 
 40~-~\citet{Burgasser02};
 41~-~\citet{Burgasser00b};
 42~-~\citet{Burgasser99}  Burgasser, A. J., Kirkpatrick, J. D., Brown, M. E., et al. 1999, ApJ, 522, L65;
 43~-~\citet{Geballe02};
 44~-~\citet{Burgasser00a};
 45~-~\citet{Burgasser03a};
 46~-~\citet{Burgasser03b};
 47~-~\citet{Wright14}
 }  
\end{deluxetable*}

\begin{deluxetable}{cccccc}
\tablewidth{80mm}
\tablecaption{Comparison with Literature Parallaxes.\label{Comp}}
\tablenum{4}
\tablehead{
\colhead{Object} & \colhead{$\pi$(mas)} &\colhead{$\pi$(mas)}  &\colhead{$\pi$(mas)}  &\colhead{$\pi$(mas)}  \\
              & \colhead{This paper} &\colhead{B14}    &\colhead{K13}    &\colhead{D\&K13} 
}
\startdata
W0148	&  91.1$\pm$2.4\ &  \nodata    & \nodata       & 60$\pm$16 \\
W0647	&  93$\pm$13     &  \nodata    & 115$\pm$12    & \nodata  \\
W0713	&  108.7$\pm$4.0 &  106$\pm$13 & \nodata       & \nodata \\
W1541	&  175.1$\pm$4.4 &  176$\pm$9  & \nodata        & 74$\pm$31 \\
W2220	&  87.2$\pm$3.7  &  106$\pm$24 & \nodata        & \nodata  \\
\enddata
\tablerefs{B14 -- \citet{Beichman14}; K13 -- \citet{Kirkpatrick13a}; D\&K13 -- \citet{DK13}.}
\end{deluxetable}

\section{Discussion}

\subsection{Spectral Type -- Absolute Magnitude Diagrams}

\begin{figure*}
   \vspace{-1.0cm}
   \figurenum{3}
   \centering\includegraphics[trim=10mm 167mm 50mm   3mm, clip=true,width=150mm]{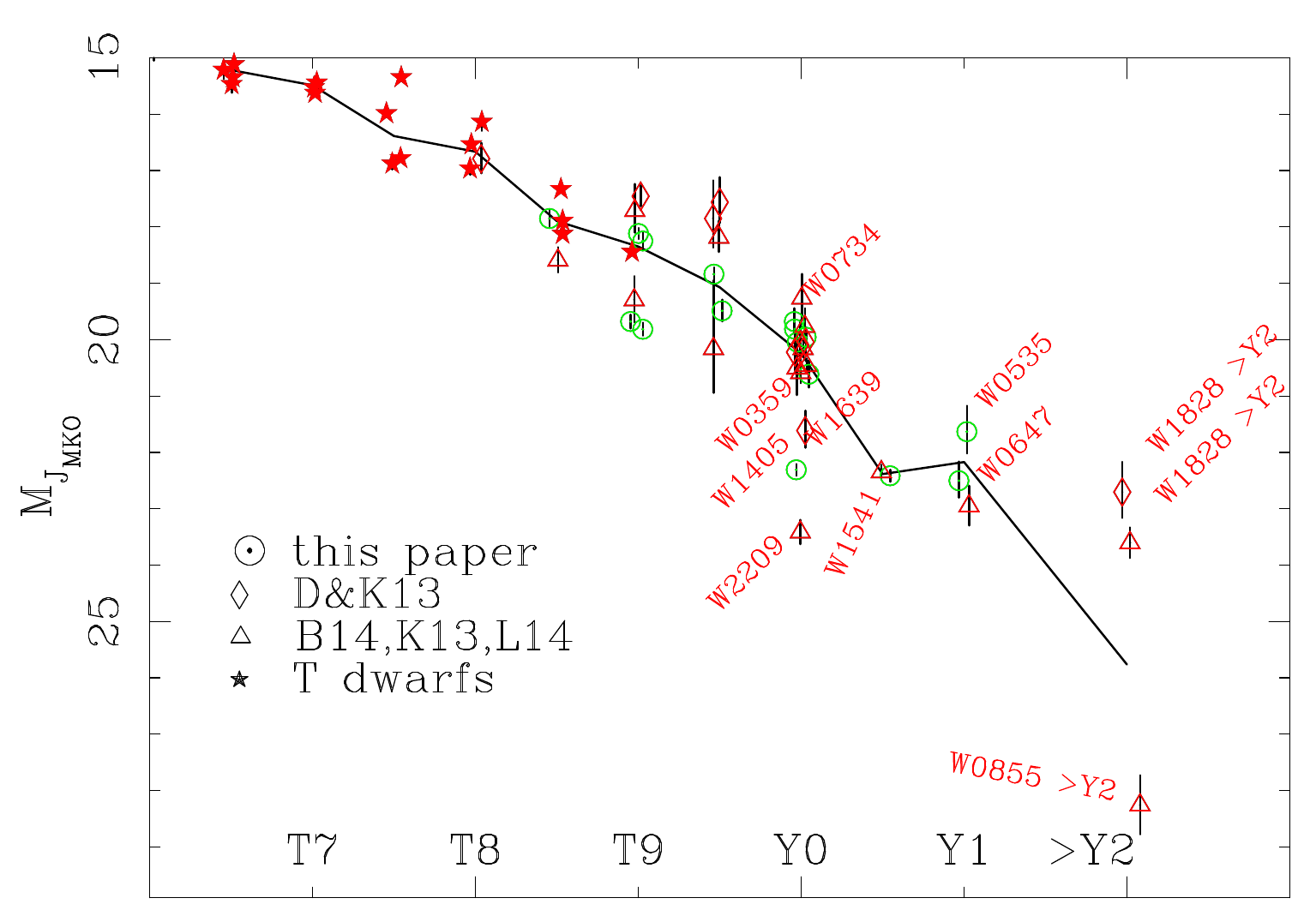}
   \centering\includegraphics[trim=10mm 11mm  50mm 146mm, clip=true,width=150mm]{Fig-1.pdf}
   \figurenum{3}
    \caption{Absolute magnitude in the J\mko\  and W2 passbands as a function of spectral types for Y dwarfs presented
            in this paper (green open circles), by \citet[D\&K13 -- red open diamonds] {DK13} and \citet[B14,K13 -- red open triangles]{Beichman14,Kirkpatrick13a}
            and for a reference sample of T dwarfs
            (red solid stars). Where multiple distances have been measured for a target, we plot each independently.
            Each data point has had a small random offset in
            spectral type applied to make the plot clearer. The solid line shows the median magnitude for all
            measurements at each spectral type. Table \ref{SpTSequence} summarises the scatter about that line at each spectral type.  
            Selected objects are highlighted where this can be done with clarity.
            The exception is W0359 which lies at the lower edge of a  cluster of Y0s near \MJ=20. This is in contrast to its isolated position
            in the colour magnitude diagrams in Figure \ref{Fig2}. 
				\label{Fig1}}
\end{figure*}

Absolute magnitudes for the Magellan objects
in J\mko\  (generated from our J3 photometry as described in \S\ref{PhotAnal}) and W2 are plotted as a function of spectral type in Figure \ref{Fig1}, 
along with the comparison data described in \S\ref{other text}.
Table \ref{SpTSequence} shows median and root-mean-square (rms) absolute magnitudes for each spectral subclass (with the exception
of W1141 which has only an estimated spectral type). 

There is substantial scatter about the median spectral type versus absolute magnitude sequences. It
has to be remembered that typical uncertainties on these spectral classifications are $\pm$0.5 sub-types --
an uncertainty that could allow W0734's over-luminous absolute magnitude to be consistent with an earlier type of T9.5
(rather than its current Y0 classification), and W1639's sub-luminous absolute magnitude to
be consistent with a later type of Y0.5 (rather than Y0). 

An unusually narrow range of luminosities has previously been reported for Y0 dwarfs by D\&K13, however the data in Fig. \ref{Fig1}
and Table \ref{SpTSequence} does not reproduce this. Our larger sample of Y0 dwarfs have a substantial rms of 1.05 and a span of
4.2 magnitudes at J. The equivalent figures at W2 are smaller, but still significant (0.32 and 1.2). 
It could be argued that a few outliers amongst the Y0 brown dwarfs
(W2209, W1405, W1639) are skewing this distribution. However we note that the removal of these three still leaves an rms of 0.44
in absolute magnitude at J and 0.23 in absolute magnitude at MW2. We return to this issue below in the context of examining
the colour magnitude diagrams. 

Finally, we note that this larger sample of
parallaxes also shows no conclusive evidence for a brightening of absolute magnitudes for T9.5 objects compared to
T9 ones as has been previously reported \citep{DK13}. 
The data are consistent (especially when the significant scatter about the median
values in each bin are considered) with objects becoming monotonically fainter from
T through Y (with the possible exception of objects later than Y0.5, as discussed further below).

\begin{deluxetable}{cccccc}
\tablewidth{80mm}
\tablenum{5}
\tablecaption{Median and rms absolute magnitudes of all independent observations of late T and Y dwarfs, 
          binned by spectral type (SpT).\label{SpTSequence}}
\tablehead{
\colhead{SpT} & \colhead{N\tablenotemark{a}}      &\colhead{M$_J$}  &\colhead{M$_J$}  &\colhead{M$_{W2}$}&\colhead{M$_{W2}$}\\
              &                                   &\colhead{median} &\colhead{rms}    &\colhead{median} &\colhead{rms} 
}
\startdata
T6.5	&  5 &  15.22 & 0.31 &    12.86 & 0.17 \\
T7.0	&  4 &  15.49 & 0.37 &    13.28 & 0.48 \\
T7.5	&  4 &  16.39 & 0.72 &    13.39 & 0.27 \\
T8.0	&  4 &  16.66 & 0.36 &    13.44 & 0.23 \\
T8.5	&  5 &  17.90 & 0.46 &    13.75 & 0.22 \\
T9.0	&  8 &  18.35 & 0.90 &    13.92 & 0.24 \\
T9.5	&  4 &  19.08 & 0.97 &    14.28 & 0.46 \\
Y0.0	& 11 &  20.32 & 1.25 &    14.65 & 0.35 \\
Y0.5	&  1 &  22.39 &\nodata&   15.20 &\nodata\\
Y1.0	&  2 &  22.18 & 0.76 &    14.78 & 0.77 \\
$>$Y2.0\tablenotemark{b}	&  2 &  25.76 & 3.52 &    15.76 & 2.15 \\
\enddata
\tablenotetext{a}{ -- N is the number of independent distance measurements in each bin. Where multiple distances are
available for a single target, they are combined in a weighted fashion before calculating these statistics. W1141, which does
not have a spectral type, is not included in these calculations.}
\tablenotetext{b}{ The $>$Y2 category contains only two members in this table and may well represent
a range in spectral types - the median and rms likely reflect this diversity.}
\end{deluxetable}

\subsection{Colour -- Absolute Magnitude Diagrams}

\begin{figure*}
   \vspace{-1.0cm}
   \figurenum{4}
   \centering\includegraphics[trim=10mm 169mm 50mm   7mm, clip=true,width=150mm]{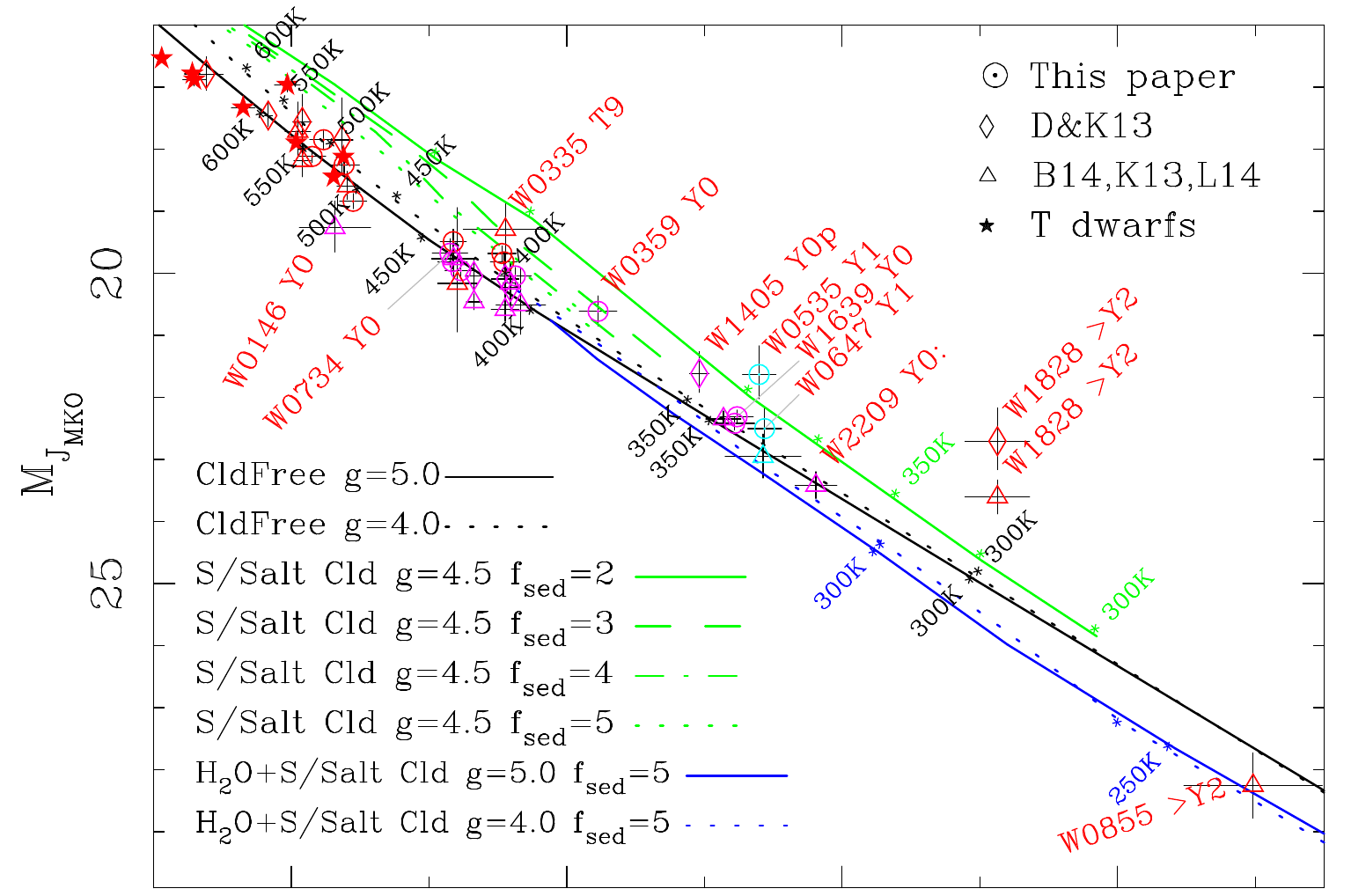}
   \centering\includegraphics[trim=10mm 10mm  50mm 149mm, clip=true,width=150mm]{Fig-2.pdf}
   \caption{Colour--Absolute Magnitude diagrams in the J\mko\  and W2 passbands. Symbols are as for Fig.\ref{Fig1} 
            except that Y0/Y0.5 objects are coloured magenta and Y1 cyan. Where multiple distances have been measured 
            for a target, we plot each independently.  Cloudless models \citep[labelled ``CldFree'', black
            lines,][]{Saumon12} are superimposed at \logg\ = 4.0 and 5.0, along with  Sulfide+Salt cloud models \citep[``S/Salt Cld'', green lines, ][]{Morley12} at
            \logg\ = 4.5 for \fsed\ = 2-5 -- with the  \fsed\ = 2 model extended by a
            special model run to \teff=300K, and water cloud models due to \citet[``H$_2$O+S/Salt Cld'', blue lines,][]{Morley14} with \logg\ = 5.0
            and 4.0 at \fsed\ = 5.   \label{Fig2}}
\end{figure*}

Spectral types do not naturally emerge from theory (i.e. from model atmospheres), as they are a fundamentally empirical classification
system driven by observed spectra. The {\em fluxes} produced by models, however, are robustly
tested by colour-magnitude diagrams. The passbands with the most complete and uniform Âdata for late-T and Y dwarfs
are the mid-infrared WISE W2 band (where all but one of the known Y dwarfs were discovered) and the near-infrared J passband (where
much WISE follow-up imaging has been done). We therefore show J\mko\  and W2 colour-magnitude diagrams in Fig. \ref{Fig2}. As
we show below these diagrams have the advantage of having minimal sensitivity to gravity.

Some objects classified as Y0 have absolute magnitudes and colours consistent with types different to those
assigned by their spectra. W0734, for example, lies at an absolute magnitude and colour consistent with being a
late T dwarf (rather than a Y dwarf), while W1639 and W2209 lie at locations suggesting photometric properties 
and effective temperatures more
in common with Y1 (and later) brown dwarfs than Y0 ones. These are  reflected in the
spectral-type-verses-absolute-magnitude diagrams presented in Fig. \ref{Fig1} (as noted above). This scatter could 
also reflect metallicity variations at the
0.3\,dex level as indicated by benchmark studies of T dwarfs \citep{Burningham13}.

The challenge posed by the molecular physics, cloud formation physics  and radiative transfer for atmospheric models
at these temperatures is considerable, and multiple generations of models have been required to reproduce observed
properties. The most straightforward are ``cloudless'' models, which assume that condensed materials drop to layers
below the photosphere and so are removed from the radiative transfer. ``CldFree'' models are shown in
Fig. \ref{Fig2} due to \citet{Saumon12} based on models by 
\citet{Morley12,SaumonMarley08} and also presented in \citet{Leggett13}. These are shown in the figure for
 surface gravities  of \logg\ = 4.0 and 5.0 \citep[the range relevant for brown dwarfs of age
100Myr-10Gyr with masses 5-30\,\mjup;][]{Baraffe03}. The impact of this plausible spread in gravities is  
small -- $\approx$0.5\,mag in \MJ\ and \MWT\ for \teff\ above 400K, dropping to less than 0.1\,mag for \teff\ at 300K and cooler.
Moreover, the impacts of gravity in both passbands are of a  similar magnitude and operate in the same direction, so that for a given J--W2
colour, there is very little sensitivity to gravity predicted by the models.

The agreement between these cloudless models and the lower envelope of observed absolute magnitudes is generally
good. Table \ref{MW2Scatter}
summarises the median and rms values (binned in J--W2) for the difference $\Delta$\MWT\ between the \logg=5.0 model and the
observed \MWT\ for each object. Given these models are a {\em prediction} (not a fit) the agreement is
striking, especially considering the complexity of the models and the history of poor matches between predictions
and observations in this field. 

That being said, some Y dwarfs are over-luminous in J\mko\  and W2 compared to these models. W0359 (Y0), W1405 (Y0)
and W0535 (Y1) sit 0.6, 0.6 and 1.1 mag (respectively) above the ``CldFree'' log(g)=5.0 model, which otherwise fits
the sequence of late T and Y dwarfs  well. W0335 is a T9 dwarf that appears similarly over-luminous (by 1.0 mag). The $>$Y2 object
W1828 is even more over-luminous. This is a long-standing issue to which we return below.


\begin{deluxetable}{ccrr}
\tablenum{6}
\tablecaption{Deviations from Cloud Free $\log g=5.0$ model}
\tablewidth{80mm}
\tablehead{
\colhead{J-W2} & \colhead{N}     &\colhead{$\Delta$M$_{W2}$\tablenotemark{a}} & \colhead{$\Delta$M$_{W2}$}\\
\colhead{range}&                 &\colhead{mean} &\colhead{rms}
}
\startdata
3.0-4.0	& 7		&	   0.12	&0.42\\
4.0-5.0	& 11	   &	   0.05	&0.35\\
5.0-6.0	& 12    &	$-$0.13	&0.40\\
6.0-7.0	& 2		&	$-$0.62	&0.01\\
7.0-8.0	& 5		&	$-$0.34	&0.49\\
8.0-11.5 & 2		&  $-$0.66	&1.60\\
\enddata
\label{MW2Scatter}
\tablenotetext{a}{Mean and rms values for the difference ($\Delta$M$_{W2}$ ) between the observed absolute magnitude and the
model W2 absolute magnitude for each object's J--W2 (in the sense $\Delta$M$_{W2}$ = ``observation'' -- ``model''). 
Where multiple distances are available for a single target, they are combined in a weighted fashion before calculating 
these statistics.}
\end{deluxetable}

No plausible range of brown dwarf radii can produce gravity variations sufficient to explain this over-luminosity.
Unresolved binarity could be the cause, since an unrecognized equal-mass binary delivers an absolute magnitude
excess of 0.75\,mag. For the distances at which these objects lie (7.5-15.6\,pc), adaptive optics imaging on 8m-class
telescopes will deliver the ability to detect binaries at separations down of much less than 1au and test this
hypothesis within a few years. 

An alternative explanation is that this over-luminosity could be due to cloud formation. Clouds are 
known to form in brown dwarfs at higher temperatures, and expected to form at the temperature ranges in question here (450-300K),
and variability suggestive of clouds has been reported in {\em Spitzer} observations for at least one Y dwarf \citep{Cushing14a}.

Cloud decks in multiple species have been predicted, with two classes of clouds predicted to be
particularly likely: sulfide and salt condensates (including Cr, MnS, Na2S, ZnS and KCl), and H$_2$O clouds. 
Models for the former
have been presented by \citet{Morley12}, by balancing the upward transport of vapor and condensate by turbulence,
with the downward transport of condensate by sedimentation, the effects of which are determined by a parameter
\fsed\  that describes the efficiency of sedimentation. High \fsed\ models have vertically thinner clouds with
larger particle sizes (i.e. optically thinner clouds), whereas low \fsed\ models have more vertically extended
clouds with smaller particles (i.e. optically thicker clouds). The predicted tracks for these ``S/Salt Cld'' models
at log(g)=4.5 for a range of \fsed\ values are also shown in Fig. \ref{Fig2} (with the \fsed=2 tracks extended to
temperatures of 300K to demonstrate their potential impact at very low temperatures).

These models do not include the effects of H$_2$O ice clouds, which will become more important at temperatures below
400K \citep{Morley12}. \citet{Morley14}  have modelled the additional impact of these clouds and they are  shown in Fig. \ref{Fig2} as
``H$_2$O+S/Salt Cld'' tracks for \fsed=5 at log(g)=4.0 and 5.0. The impact of these two types of cloud are quite
distinct -- sulphide and salt clouds make models at equivalent effective temperatures redder in J--W2, so that the
models appear to sit at higher luminosities in Fig. \ref{Fig2} compared to Cloudless tracks. H$_2$O clouds, on the other hand,
make models at equivalent effective temperatures bluer in J--W2, so that these model tracks appear to sit at lower
luminosities than Cloudless models. 

The sulphide and salt clouds redden J--W2 because they reside deep in the atmosphere. The J band lies in a window between 
molecular absorption features in which we usually see very deeply, however clouds limit the depth to which we observe, 
decreasing the J band flux. The W2 flux emerges from higher altitudes, well above the salt/sulfide clouds, so W2 changes very little. 
In comparison, the water ice clouds are formed higher in the atmosphere where the mid-infrared flux emerges. 
In addition, the optical properties of water ice mean that it absorbs much more strongly in the mid-infrared than in the optical-through-J 
band (where it scatters but absorbs little). This means that we see a similar flux emerging in the J band, but a decreased  flux in W2, 
making the models appear bluer in J--W2.

In the main, most of the Y dwarfs with distance measurements are consistent
with cloud-free model predictions. Nonetheless a few Y dwarfs display an over-luminosity in the near-infrared that
could be caused (assuming they are not shown to be binaries) by the presence of thick clouds of sulphide and salt
condensates at temperatures of 400K or below. 
In contrast, W0855 \citep{Luhman14} lies in a region of the colour magnitude diagram (extremely red in J--W2 and
sub-luminous for its colour in \MWT) requiring the presence of water ice clouds,
making it the first system to display this behaviour \citep{Faherty14b}. 

None of the models
predict the over-luminosity of W1828, which remains an enigma. In particular, the suggestion that this system could be a
325K+300K binary \citep{Leggett13} seems hard to reconcile with it being observed to lie 2 magnitudes above the Cloudless
model sequence in \MWT\ (Fig. \ref{Fig2}), 2.5 magnitudes brighter than the H$_2$O Cloudy model sequence, and 3.2 magnitudes brighter
than W0855 (which is otherwise consistent with the H$_2$O Cloudy model sequence).
For W1828's over luminosity relative
to both W0855 and these models to be explicable by binarity, it would have to be an nearly-equal-mass-triple or -quadruple system.
Between them W1828 and W0855 certainly indicate that even the coldest brown dwarfs currently known display a 
large spread in absolute magnitude for a given colour.\footnote{Though we note  that these models do not yet 
include the effects of metallicity which will certainly have some impact
 (along with disequilibrium chemistry) in the emergent spectra of these Y dwarfs.}

The fact that some Y dwarfs are significantly over-luminous (for a given colour), 
while others are not, suggests the extent of
cloud coverage could vary significantly between otherwise similar Y dwarfs. This would either imply that cloud
{\em thickness}, or that {\em cloud coverage} varies. The latter would mean that cloud sensitive photometric or
spectroscopic features can be expected to vary as these objects rotate. Precisely this type of variability has been
seen at the L-T spectral type transition \citep{Radigan12,Artigau09,Crossfield14}, for mid-T dwarfs \citep{Buenzli14},
and is predicted in late T dwarfs \citep{Morley12} due to
sulphide and salt clouds. Recent {\em Spitzer} results indicate some Y dwarfs are
indeed variable on timescales suggesting the presence of clouds \citep{Cushing14a}.
It will be important (despite the severe technical challenges posed by these faint
targets) to target Y dwarf variability as a probe of their cloud structures and properties.

\acknowledgements

This paper includes data gathered with the 6.5 meter Magellan Telescopes located at Las Campanas Observatory, Chile. Australian access to the Magellan
Telescopes was supported through the National Collaborative Research Infrastructure and Collaborative Research Infrastructure Strategies of the
Australian Federal Government. Access through the Chilean Time Allocation Committee was supported by awards CN2012A-011, CN2012B-057, CN2013A-127.
This research was supported by Australian Research Council grants DP0774000 and DP130102695.
This publication makes use of data products from the Wide-field Infrared Survey Explorer (WISE), which is a joint project of the University of
California, Los Angeles, and the Jet Propulsion Laboratory/California Institute of Technology, and NEOWISE, which is a project of the Jet Propulsion
Laboratory/California Institute of Technology. WISE and NEOWISE are funded by the US National Aeronautics and Space Administration.\\

{\it Facilities:} \facility{Magellan:Baade (FourStar)}, \facility{WISE}

\clearpage
\LongTables
\begin{landscape}
\begin{deluxetable*}{llllllcccclccc}
\tablenum{2}
\tabletypesize{\footnotesize}
\tablecaption{Photometric \& Astrometric Data for Magellan Targets\label{Results}}
\tablewidth{0pt}
\tablehead{
\multicolumn{2}{l}{Designations} &\colhead{SpT}  &\colhead{J3}   &\colhead{J\mko}&\colhead{W2}   &\colhead{$\pi$}&\colhead{$\mu$}  &\colhead{$\theta$}&\colhead{V$_\mathrm{tan}$}&\colhead{N$_\mathrm{epoch}$,}&\colhead{Disc} & \colhead{Phot} \\
\colhead{Short}   & \colhead{\em WISE}  &               &\colhead{(mag)}&\colhead{(mag)}&\colhead{(mag)}&\colhead{(mas)}&\colhead{(mas/y)}&\colhead{($\deg$)}&\colhead{km/s}            &\colhead{N$_\mathrm{ref}$}   &\colhead{Ref} & \colhead{Ref} 
}

\startdata
W0148	& WISEA J014807.34-720258.7  &T9.5  &  18.83$\pm$0.02 &  18.96$\pm$0.07 &  14.592$\pm$0.039 & 91.1$\pm$3.4  &   1269.3$\pm$4.1 &  90.0$\pm$0.2 &  66$\pm$13    & 7,18 & 2,6 & 2 \\
W0359	& WISEA J035934.07-540154.8  &Y0    &  21.40$\pm$0.09 &  21.56$\pm$0.24 &  15.384$\pm$0.054 & 63.2$\pm$6.0  &   765$\pm$12     & 193.3$\pm$0.8 &  57$\pm$18    & 7,8  & 4 & 4 \\
W0535	& WISEA J053516.87-750024.6  &Y1    &  22.09$\pm$0.07 &  \nodata        &  14.904$\pm$0.047 & 74$\pm$14     &   119$\pm$16     & 287.7$\pm$3.7 &  7.7$\pm$3.5  & 5,21 & 4 &   \\
W0647	& WISEA J064723.24-623235.4  &Y1    &  22.45$\pm$0.07 &  22.65$\pm$0.27 &  15.224$\pm$0.051 & 93$\pm$13     &   368$\pm$18     &   0.1$\pm$2.5 &  18.8$\pm$7.0 & 4,37 & 5 & 5 \\
W0713	& WISEA J071322.55-291752.0  &Y0    &  19.42$\pm$0.03 &  19.64$\pm$0.15 &  14.462$\pm$0.052 & 108.7$\pm$4.0 &   540.2$\pm$7.3  & 139.6$\pm$0.1 &  23.6$\pm$4.6 & 7,21 & 4 & 3 \\
W0734	& WISEA J073444.03-715743.8  &Y0    &  20.13$\pm$0.08 &  20.41$\pm$0.27 &  15.189$\pm$0.050 & 73.7$\pm$6.6  &   571.6$\pm$7.7  & 261.8$\pm$0.8 &  37$\pm$11    & 7,32 & 4 & 4 \\
W0811	& WISEA J081117.95-805141.4  &T9.5  &  19.31$\pm$0.01 &  \nodata        &  14.345$\pm$0.038 & 98.5$\pm$7.7  &   293.4$\pm$6.9  & 100.6$\pm$0.9 &  14.1$\pm$4.0 & 6,19 & 7 &   \\
W1042	& WISEA J104245.24-384238.1  &T8.5  &  18.58$\pm$0.02 &  \nodata        &  14.556$\pm$0.048 & 64.8$\pm$3.4  &   93.7$\pm$6.2   & 143.3$\pm$1.4 &  6.9$\pm$1.6  & 9,26 & 6 &   \\
W1141	& WISEA J114156.67-332635.5  &(Y0)\tablenotemark{a}
                                         &  19.63$\pm$0.05 &  19.76$\pm$0.14 &  14.611$\pm$0.055 & 105.5$\pm$4.3 &   901.0$\pm$6.3  & 264.6$\pm$0.4 &  40.5$\pm$8.2 & 7,20 & 3 & 3 \\
W1541	& WISEP J154151.65-225025.2\tablenotemark{b}
                                  &Y0.5  &  20.99$\pm$0.03 &  21.12$\pm$0.06 &  13.982$\pm$0.112 & 175.1$\pm$4.4 &   899.0$\pm$4.2  & 264.4$\pm$0.3 &  24.3$\pm$3.9 & 8,23 & 6 & 1 \\
W1639	& WISEA J163940.84-684739.4  &Y0    &  20.57$\pm$0.05 &  \nodata        &  13.544$\pm$0.059 & 202.3$\pm$3.1 &   3156.0$\pm$3.5 & 169.3$\pm$0.1 &  73.9$\pm$9.2 & 8,13 & 8 &   \\
W2102	& WISEA J210200.14-442919.9  &T9    &  18.08$\pm$0.01 &  18.24$\pm$0.04 &  14.139$\pm$0.043 & 92.3$\pm$1.9  &   356.9$\pm$2.7  & 173.3$\pm$0.3 &  18.3$\pm$2.6 & 8,18 & 4 & 3 \\
W2134	& WISEA J213456.79-713744.7  &T9p   &  19.28$\pm$0.04 &  \nodata        &  13.962$\pm$0.036 & 109.1$\pm$3.7 &   1381.4$\pm$6.2 &  99.3$\pm$0.2 &  60$\pm$11    & 6,21 & 2 &   \\
W2220	& WISEA J222055.34-362817.5  &Y0    &  20.13$\pm$0.02 &  20.47$\pm$0.11 &  14.714$\pm$0.056 & 87.2$\pm$3.7  &   297.9$\pm$5.2  & 108.4$\pm$0.5 &  16.2$\pm$3.3 & 9,22 & 2 & 3 \\
W2325	& WISEA J232519.55-410535.1  &T9p   &  19.44$\pm$0.02 &  19.75$\pm$0.05 &  14.108$\pm$0.040 & 107.8$\pm$3.7 &   837.0$\pm$6.7  &  91.2$\pm$0.6 &  36.8$\pm$6.8 & 5,16 & 2 & 2 \\
W2332	& WISEA J233226.54-432510.9  &T9    &  19.13$\pm$0.02 &  19.40$\pm$0.10 &  14.958$\pm$0.066 & 60.5$\pm$4.0  &   355.6$\pm$7.4  & 136.0$\pm$0.1 &  27.9$\pm$7.2 & 6,22 & 4 & 3 \\
\enddata
\tablerefs{1~-~\citet{Leggett13}, 2~-~\citet{kirkpatrick2011}, 3~-~C.G. Tinney et al. 2014, in preparation,  4~-~\citet{kirkpatrick2012}, 5~-~\citet{Kirkpatrick13a}, 
                                                              6~-~\citet{Cushing11}, 7~-~\citet{Mace13}, 8~-~\citet{Tinney12}}
\tablenotetext{a}{The spectral type for W1141 has been estimated from its photometry as described in the text.}
\tablenotetext{b}{The parallax solution for W1541 is based on the right ascension solution alone as described in the text.}
\end{deluxetable*}

\clearpage

\end{landscape}


\begin{thebibliography}{} 
\bibitem[Allard et al. (2003)]{Allard03}                  Allard, F., Guillot, T., Ludwig, H.-G., et al. 2003, in IAU Symposium 211, Brown Dwarfs, ed. E. Martõn (San Francisco, CA: ASP), 325                 
\bibitem[Artigau et al. (2009)]{Artigau09}                Artigau, E., Bouchard, S., Doyon, R., \& Lafreniere, D. 2009, ApJ, 701, 1534                                                                          
\bibitem[Baraffe et al. (2003)]{Baraffe03}                Baraffe, I., Chabrier, G., Barman, T., Allard, F., Hauschildt, P. 2003, A\&A, 402, 701                                                                
\bibitem[Beichman et al. (2014)]{Beichman14}              Beichman, C., Gelino, C. R., Kirkpatrick, J. D., Cushing, M. C., Dodson-Robinson, S., Marley, M.S., Morley, C. V., Wright, E. L., 2014, ApJ, 783, 68 (B13)
\bibitem[Bertin \& Arnouts (1996)]{bertin96}              Bertin, E. \&  Arnouts, S., A\&AS, 117, 393
\bibitem[Buenzli et al. (2014)]{Buenzli14}                Buenzli, E., Apai, D., Radigan, J., Reid, I. N., Flateau, D., 2014, ApJ, 782, 77                                                                     
\bibitem[Burgasser et al. (1999)]{Burgasser99}            Burgasser, A. J., Kirkpatrick, J. D., Brown, M. E., et al. 1999, ApJ, 522, L65
\bibitem[Burgasser et al. (2002)]{Burgasser02}            Burgasser, A. J., Kirkpatrick, J. D., Brown, M. E., et al. 2002, ApJ, 564, 421
\bibitem[Burgasser et al. (2006)]{Burgasser06}				 Burgasser, A. J., Kirkpatrick, J. D., Cruz, K. L., Reid, I. N., Leggett, S. K., Liebert, J., Burrows, A., Brown, M. E. 2006, ApJS, 166, 585
\bibitem[Burgasser et al. (2000a)]{Burgasser00a}          Burgasser, A. J., Kirkpatrick, J. D., Cutri, R. M., et al. 2000a, ApJ, 531, L57
\bibitem[Burgasser et al. (2003a)]{Burgasser03a}          Burgasser, A. J., Kirkpatrick, J. D., McElwain, M. W., et al. 2003a, AJ, 125, 850
\bibitem[Burgasser et al. (2003b)]{Burgasser03b}          Burgasser, A. J., McElwain, M. W., Kirkpatrick, J. D. 2003b, AJ, 126, 2487
\bibitem[Burgasser et al. (2004)]{Burgasser04}            Burgasser, A. J., McElwain, M. W., Kirkpatrick, J. D., et al. 2004, AJ, 127, 2856
\bibitem[Burgasser et al. (2008)]{Burgasser08}				 Burgasser, A. J., Tinney, C. G., Cushing, M. C., et al. 2008, 689, L53
\bibitem[Burgasser et al. (2000b)]{Burgasser00b}          Burgasser, A. J., Wilson, J. C., Kirkpatrick, J. D., et al. 2000b, AJ, 120, 1100
\bibitem[Burningham et al. (2008)]{Burningham08}		    Burningham, B., Pinfield, D. J., Leggett, S. K. et al. 2008, MNRAS, 391, 320
\bibitem[Burningham et al. (2013)]{Burningham13}          Burningham, B. et al. 2013, MNRAS, 433, 457                                                                                                                                       
\bibitem[Chiu et al. (2006)]{Chiu06}						    Chiu, K., Fan, X., Leggett, S. K., Golimowski, D. A., Zheng, W., Geballe, T. R., Schneider, D. P., Brinkmann, J., 2006, ApJ, 131, 2722
\bibitem[Crossfield et al. (2014)]{Crossfield14}          Crossfield, I.J.M et al. 2014, Nature, 505, 654                                                                                                                                   
\bibitem[Cushing et al. (2011)]{Cushing11}                Cushing, M.~C., Kirkpatrick, J.~D., Gelino, C.~R., et al.\ 2011, \apj, 743, 50                                                                                                     
\bibitem[Cushing et al. (2014a)]{Cushing14a}      		    Cushing, M., Hardegree-Ullman, K., Trucks, J, 2014a, AAS, 223, 425.08
\bibitem[Cushing et al. (2014b)]{Cushing14b}      		    Cushing, M. C., Kirkpatrick, J. D., Gelino, C. R., Mace, G. N., Skrutskie, M. F., Gould, A., 2014b, AJ, 147, 113
\bibitem[Dahn et al. (2002)]{Dahn02}                      Dahn, C. C., Harris, H. C., Vrba, F. J. et al. 2002, AJ, 124, 1170                                                                                                                                            
\bibitem[Delorme et al. ()2008]{Delorme08}				    Delorme, P., Delfosse, X., Albert, L. et al. 2008, A\&A, 482, 961
\bibitem[Dupuy \& Kraus (2013)]{DK13}                     Dupuy, T. J. \& Kraus, A. L. 2013, 2013, 341, 1492 (DK13)                                                                                                                              
\bibitem[Dupuy \& Liu (2012)]{DL12}                       Dupuy, T. J. \& Liu, M. C. 2012, ApJS, 201, 19                                                                                                                                     
\bibitem[Faherty et al. (2012)]{Faherty12}				    Faherty, J. K., Burgasser, A. J., Walter, F. M. et al., 2012, ApJ, 752, 56
\bibitem[Faherty et al. (2014)]{Faherty14b}				    Faherty, J. K., Tinney, C. G., Monson, A. et al., 2014, ApJL, 793, L16
\bibitem[Geballe et al. (2001)]{Geballe01}				    Geballe, T. R., Saumon, D., Leggett, S. K., Knapp, G. R. ,Marley, M. S., Lodders, K., 2001, ApJ, 556, 373
\bibitem[Geballe et al. (2002)]{Geballe02}				    Geballe, T. R., Knapp, G. R., Leggett, S. K., et al. 2002, ApJ, 564, 466
\bibitem[Kirkpatrick et al. (2011)]{kirkpatrick2011}      Kirkpatrick, J.~D., Cushing, M.~C., Gelino, C.~R., et al.\ 2011, \apjs, 197, 19                                                                                                    
\bibitem[Kirkpatrick et al. (2012)]{kirkpatrick2012}      Kirkpatrick, J.~D., Gelino, C.~R., Cushing, M.~C., , et al.\ 2012, \apj, 753, 156                                                                                                  
\bibitem[Kirkpatrick et al. (2013)]{Kirkpatrick13a}      Kirkpatrick, J. D., Cushing, M. C., Gelino, C. R. et al. 2013, ApJ, 776, 128  (K13)                                                                                                                                    
\bibitem[Knapp et al. (2004)]{Knapp04}						    Knapp, G. R., Leggett, S. K., Fan, X. et al., 2004, ApJ, 127, 3553
\bibitem[Leggett et al. (2010)]{Leggett10}				    Leggett, S. K., Burningham, B., Saumon, D. 2010, ApJ, 710, 1627
\bibitem[Leggett et al. (2009)]{Leggett09}				    Leggett, S. K., Cushing, M. C., Saumon, D. et al. 2009, ApJ, 695, 1517
\bibitem[Leggett et al. (2002)]{Leggett02}				    Leggett, S. K., Golimowski, D. A., Fan, X. et al. 2002, ApJ, 564, 452
\bibitem[Leggett et al. (2012)]{Leggett12}				    Leggett, S. K., Saumon, D., Marley, M. S., et al. 2012, ApJ, 748, 74 
\bibitem[Leggett et al. (2013)]{Leggett13}                Leggett, S. K., Morley, C.V., Marley, M.S., Saumon, D., Fortney, S.J., Visscher, C., 2013, ApJ, 763, 130                                                                            
\bibitem[Liu et al. (2011a)]{Liu11}							    Liu, M. C., Deacon, Niall R., Magnier, E. A. et al.  2011a, ApJ, 740, L32
\bibitem[Liu et al. (2011b)]{Liu11b}							    Liu, M. C., Delorme, P. Dupuy, T.J. et al.  2011b, ApJ, 740, 108
\bibitem[Lodieu et al. (2007)]{Lodieu07}					    Lodieu, N., Pinfield, D. J., Leggett, S. K. et al., 2007,  MNRAS, 379, 1423
\bibitem[Looper et al. (2007)]{Looper07}                  Looper, D. L., Kirkpatrick, J. D., \& Burgasser, A. J. 2007, AJ, 134, 1162                                                                                                         
\bibitem[Lucas et al. (2010)]{Lucas10}                    Lucas, P. W., Tinney, C. G., Burningham, B. et al. 2010, MNRAS, 408, L56                                                                                                           
\bibitem[Luhman (2014)]{Luhman14}                         Luhman, K.L., 2014, ApJ, 786, L18  (L14)                                                                                                                                                
\bibitem[Luhman et al. (2011)]{Luhman11}                  Luhman, K.L., Burgasser, A. J., Bochanski, J. J., 2011, ApJ, 730, L9                                                                                                                     
\bibitem[Mace et al. (2013)]{Mace13}                      Mace, G. N., Kirkpatrick, J. D., Cushing, M. C., et al. 2013, ApJS, 205, 6                                                                                                         
\bibitem[Mainzer et al. (2011)]{Mainzer11}                Mainzer, A., Bauer, J., Grav, T. et al. 2011, ApJ, 731, 53                                                                                                                                             
\bibitem[Marocco F. et al. (2010)]{Marocco10}             Marocco F., Smart R.L., Jones H.R.A., et al. 2010, A\&A, 524, 38                                                                                                                   
\bibitem[Marsh et al. (2013)]{Marsh13}                    Marsh, K. A. et al. 2013, ApJ, 762, 119                                                                                                                                            
\bibitem[Monet et al. (1992)]{Monet92}                    Monet, D.G., Dahn, C.C., Vrba, F.J., Harris, H.C., Pier, J.R., Luginbuhl, C.B., Ables, H.D., 1992 AJ, 103, 638                                                                     
\bibitem[Morley et al. (2012)]{Morley12}                  Morley, C. V., Fortney, J. J., Marley, M. S., Visscher, C., Saumon, D. \& Leggett, S. K. 2012, ApJ, 756, 172                                                                       
\bibitem[Morley et al. (2014)]{Morley14}                  Morley, C. V., Marley, M. S., Fortney, J. J. et al. 2014, ApJ, 787, 78                                                                                                                      
\bibitem[Persson et al. (2008)]{persson2008}              Persson, S. E., Barkhouser, R., Birk, C., et al. 2008, Proc. SPIE, 7014, 95                                                                                                        
\bibitem[Radigan et al. (2012)]{Radigan12}                Radigan, J., Jayawardhana, R., Lafreniere, D., Artigau, E., Marley, M., Saumon, D., 2012, ApJ, 750, 105                                                                            
\bibitem[Reid \& Hawley (2006)]{ReidHawley06}             Reid, I.N.R. \& Hawley, S.L. 2006, New Light on Dark Stars (Springer:Heidelberg)                                                                                                   
\bibitem[Saumon \& Marley (2008)]{SaumonMarley08}         Saumon. D. \& Marley M. S. 2008, ApJ, 689, 1327                                                                                                                                    
\bibitem[Saumon et al. (2012)]{Saumon12}					     Saumon. D., Marley, M. S., Abel, M., Frommhold, L., Freedman, R. S., 2012, ApJ, 750, 74                                                                                                      
\bibitem[Scholz et al. (2011)]{Scholz11}                  Scholz, R.-D., Bihain, G., Schnurr, O., \& Storm, J. 2011, A\&A, 532, L5                                                                                                            
\bibitem[Skrutskie et al. (2006)]{skrutskie2006}          Skrutskie, M.~F., Cutri, R.~M., Stiening, R., et al.\ 2006, \aj, 131, 1163                                                                                                         
\bibitem[Smart (2014)]{smart14}                           Smart, R. L., Mem. S. A. It. 2014, in press                                                                                                         
\bibitem[Stetson (1987)]{Stetson87}                       Stetson, P.B., 1987, PASP, 99, 191                                                                                                                                                 
\bibitem[Stevenson (1991)]{Stevenson91}                   Stevenson, D. 1991, ARA\&A, 29, 163                                                                                                                                                
\bibitem[Strauss M.A. et al. (1999)]{Strauss99}           Strauss M.A., Fan X., Junn J.E. et al. 1999, ApJ, 522, 61                                                                                                                          
\bibitem[Tinney et al. (2003)]{Tinney03}                  Tinney, C. G., Burgasser, A. J., \& Kirkpatrick, J. D. 2003, AJ, 126, 975                                                                                                          
\bibitem[Tinney et al. (2005)]{Tinney05}                  Tinney, C. G., Burgasser, A. J., Kirkpatrick, J. D., McElwain, M. W., 2005, AJ, 130, 2326                                                                                    
\bibitem[Tinney et al. (1995)]{Tinney95}                  Tinney, C. G., Reid, I. N., Gizis, J., \& Mould, J. R. 1995, AJ, 110, 3014                                                                                                         
\bibitem[Tinney et al. (2012)]{Tinney12}                  Tinney, C. G., Faherty, J. K., Kirkpatrick, J. D. et al. 2012, ApJ, 759, 60                                                                                                                                             
\bibitem[Tsuji (2000)]{Tsuji00}                           Tsuji T. 2000. In Very Low-Mass Stars and Brown Dwarfs, ed. R Rebolo, MR Zapatero Osorio, pp. 156--68. Cambridge, UK: Cambridge Univ. Press                                        
\bibitem[Tsvetanov et al. (2000)]{Tsvetanov00}            Tsvetanov, Z. I., Golimowski, D. A., Zheng, Wei, 2000, ApJ, 531, 61                                                                                                                
\bibitem[Vrba et al. (2004))]{Vrba04}                     Vrba, F. J. et al., 2004, AJ, 127, 2948                                                                                                                                            
\bibitem[Warren et al. (2007)]{Warren07}                  Warren, S. J., Mortlock, D. J., Leggett, S. K. et al. 2007,  MNRAS, 381, 1400                                                                                                      
\bibitem[Wright et al. (2010)]{Wright10}                  Wright, E.L. et al.  2010 AJ, 140, 1868. 
\bibitem[Wright et al. (2014)]{Wright14}                  Wright, E. L., Mainzer, A., Kirkpatrick, J. D., et al. 2014, arXiv:1405.7350
\bibitem[van Leeuwen et al. (2007)]{van Leeuwen07}        van Leeuwen, F. 2007, A\&A, 474, 653                                                                                                                                               

\end{thebibliography}
\end{document}